\documentclass[sigconf]{acmart}

\usepackage{amsmath,amsfonts}
\usepackage[ruled, linesnumbered, noend]{algorithm2e}
\usepackage{graphicx}
\usepackage{textcomp}
\usepackage{multirow}
\usepackage{float}
\usepackage{hyperref}
\usepackage{pifont}
\usepackage{amsmath}

\DeclareMathOperator*{\argmin}{arg\,min}

\renewcommand{\arraystretch}{0.5} 

\newcommand{\vspacesections}{\vspace{-0.6mm}}
\newcommand{\mysection}[1]{\vspacesections \section{#1} \vspacesections}
\newcommand{\mysubsection}[1]{\vspacesections \subsection{#1} \vspacesections}
\newcommand{\mysubsubsection}[1]{\vspacesections \subsubsection{#1} \vspacesections}

\def \circle[#1]{\raisebox{.2pt}{\textcircled{\raisebox{-.9pt} {#1}}}}
\def \square[#1]{\fbox{\textbf{{#1}}}}
\def \numberlabel[#1]{\textbf{({#1})}}

\newcommand{\weightslice}{weight slice}
\newcommand{\inputslice}{input slice}
\newcommand{\columnsum}{column sum}
\newcommand{\slicedproduct}{sliced product}

\newcommand{\Columnsum}{Column sum}

\newcommand{\IronLaw}{Titanium Law}
\newcommand{\WeightCountLimited}{Weight-Count-Limited}
\newcommand{\SumFidelityLimited}{Sum-Fidelity-Limited}
\newcommand{\ie}{{\em i.e.}, }
\newcommand{\eg}{{\em e.g.}, }
\makeatletter
\def\blfootnote{\gdef\@thefnmark{}\@footnotetext}
\makeatother

\copyrightyear{2023}
\acmYear{2023}
\setcopyright{rightsretained}
\acmConference[ISCA '23] {Proceedings of the 50th Annual International Symposium on Computer Architecture}{June 17--21, 2023}{Orlando, FL, USA.}
\acmBooktitle{Proceedings of the 50th Annual International Symposium on Computer Architecture (ISCA '23), June 17--21, 2023, Orlando, FL, USA}
\acmPrice{}
\acmISBN{979-8-4007-0095-8/23/06}
\acmDOI{10.1145/3579371.3589062}

\begin{document}

\title{RAELLA: Reforming the Arithmetic for Efficient, Low-Resolution, and Low-Loss Analog PIM: No Retraining Required!}


\author{Tanner Andrulis}
\affiliation{%
  \institution{Massachusetts Institute of Technology}
  \streetaddress{77 Massachusetts Avenue}
  \city{Cambridge}
  \state{Massachusetts}
  \country{USA}
  \postcode{02139}
}
\email{andrulis@mit.edu}

\author{Joel S. Emer}
\affiliation{%
  \institution{Massachusetts Institute of Technology, Nvidia}
  \streetaddress{77 Massachusetts Avenue}
  \city{Cambridge}
  \state{Massachusetts}
  \country{USA}
  \postcode{02139}
}
\email{jsemer@mit.edu}

\author{Vivienne Sze}
\affiliation{%
  \institution{Massachusetts Institute of Technology}
  \streetaddress{77 Massachusetts Avenue}
  \city{Cambridge}
  \state{Massachusetts}
  \country{USA}
  \postcode{02139}
}
\email{sze@mit.edu}

\begin{abstract}
    Processing-In-Memory (PIM) accelerators have the potential to efficiently run Deep Neural Network (DNN) inference by reducing costly data movement and by using resistive RAM (ReRAM) for efficient analog compute. Unfortunately, overall PIM accelerator efficiency is limited by energy-intensive analog-to-digital converters (ADCs). Furthermore, existing accelerators that reduce ADC cost do so by changing DNN weights or by using low-resolution ADCs that reduce output fidelity. These strategies harm DNN accuracy and/or require costly DNN retraining to compensate.

    To address these issues, we propose the RAELLA architecture. RAELLA adapts the architecture to each DNN; it lowers the resolution of computed analog values by encoding weights to produce near-zero analog values, adaptively slicing weights for each DNN layer, and dynamically slicing inputs through speculation and recovery. Low-resolution analog values allow RAELLA to both use efficient low-resolution ADCs and maintain accuracy without retraining, all while computing with fewer ADC converts.
    
    Compared to other low-accuracy-loss PIM accelerators, RAELLA increases energy efficiency by up to 4.9$\times$ and throughput by up to 3.3$\times$. Compared to PIM accelerators that cause accuracy loss and retrain DNNs to recover, RAELLA achieves similar efficiency and throughput without expensive DNN retraining.
\end{abstract}




\received{22 November 2022}
\received[revised]{21 February 2023}
\received[accepted]{9 March 2023}

\maketitle

\mysection{Introduction}
    \blfootnote{Models of RAELLA are available at https://github.com/mit-emze/raella}
    Processing-In-Memory (PIM) is a promising solution to the high compute and energy cost of Deep Neural Network (DNN) inference. By computing in memory~\cite{reram_survey}, PIM accelerators avoid expensive off-chip movement of the DNN weights~\cite{efficient_processing_of_dnns}. Furthermore, PIM accelerators often utilize Resistive-RAM (ReRAM) devices and ReRAM crossbars~\cite{ISAAC, PRIME, PipeLayer} for dense and efficient analog compute~\cite{reram_survey}.
    
    Unfortunately, while ReRAM crossbars can compute efficiently and with high density, overall PIM accelerator energy is often dominated by the analog-to-digital converters (ADCs) that read computed analog values from crossbars. Due to ADC overhead, some PIM accelerators~\cite{ISAAC, AtomLayer} do not significantly improve energy over non-PIM accelerators~\cite{DaDianNao} despite the opportunities in PIM.
    
    Some prior works attempt to reduce this ADC overhead by reducing the resolution of the ADC, which exponentially decreases ADC energy~\cite{ADC_Scaling_Murmann}. Architectures often partition, or \emph{slice}, the bits in DNN inputs and weights into multiple lower-resolution slices and compute with different slices in multiple steps~\cite{ISAAC}. Although sliced arithmetic can use lower-resolution ADCs, ADCs must process the results of each slice, so \emph{these strategies replace each high-resolution ADC convert with multiple low-resolution ADC converts}, and therefore ADCs still dominate overall energy. 
    
    Other PIM accelerators reduce ADC energy, but do so at the expense of DNN accuracy. Some designs prune DNNs~\cite{FORMS,SRE,ASBP,PIM-Prune,learning_sparsity_for_reram} to reduce DNN weight count, so we call these designs \emph{\WeightCountLimited{}}. They reduce the computation count and ADC converts required, but also introduce accuracy loss. Alternatively, other designs use efficient lower-resolution ADCs to process high-resolution analog values from crossbars~\cite{TIMELY,PRIME,CASCADE}. We call these designs \emph{\SumFidelityLimited{}} as the resolution difference reduces output fidelity and introduces error. These architectures requantize DNNs to tolerate ADC resolution limitations, which again causes accuracy loss.
    
    To reduce this accuracy loss, both \WeightCountLimited{} and \SumFidelityLimited{} architectures retrain DNNs. This is a problem; DNN training has a very high computational cost~\cite{DNN_energy}, can require cumbersome hyperparameter tuning to achieve high accuracy~\cite{hyperparameter_tuning}, and may be impossible if the training data is private~\cite{deepface_proprietary, dalle_proprietary}. Furthermore, cutting-edge DNNs often require particular training schemes~\cite{ibm_low_precision_inference}, which may not be compatible with the retraining scheme required by an architecture.
    
    To avoid accuracy loss without imposing retraining, we look at fidelity limitations. We define \emph{fidelity} as the ability of the ADC to represent the full resolution of computed analog values. Architectures lose fidelity and generate errors when the computed analog value resolution is higher than the ADC resolution. Each DNN produces many distributions of analog values, and prior \SumFidelityLimited{} approaches modify DNNs to reshape these analog value distributions to fit a resolution-limited ADC range. In contrast, we observe that we can reshape analog value distributions with an adaptable architecture, rather than changing the DNN.
    
    Using this key insight, we propose the RAELLA architecture to enable efficient PIM inference without retraining. RAELLA modifies arithmetic and slicing, shaping computed value distributions to produce low-resolution analog results. This allows RAELLA to use efficient low-resolution ADCs while maintaining high fidelity and low DNN accuracy loss. The main contributions of RAELLA are:

    \begin{itemize}
    \item Center+Offset encoding to accumulate more values in the analog domain while keeping small, low-resolution sums. Specifically, RAELLA shifts DNN weights to equalize the average magnitude of the positive and negative weight slices in each crossbar column. As analog-domain calculations are accumulated, positive and negative results negate to produce near-zero sums that can be converted with high fidelity.
    \item Adaptive Slicing of DNN weights at compilation time to balance density, efficiency, and fidelity. Storing more bits in each ReRAM device is denser and more efficient but creates higher-resolution analog values. For each DNN layer, RAELLA adapts the number of ReRAM devices per weight and the number of bits in each ReRAM device. This enables RAELLA to use the densest and most efficient strategies possible while keeping computed analog values low-resolution.
    \item Dynamic Slicing of DNN input activations at runtime for both efficient and high-fidelity computation. RAELLA speculates with an efficient strategy that processes with more bits in each input slice. RAELLA detects and recovers from incorrect results using a less efficient, higher-fidelity strategy that processes inputs with more slices using fewer bits each. This allows RAELLA to further reduce the number of ADC conversions without reducing fidelity.
    \end{itemize}
    
    Compared to other low-accuracy-loss PIM accelerators~\cite{ISAAC}, RAELLA can both lower ADC resolution and run DNNs with up to $14\times$ fewer ADC conversions without sacrificing fidelity.
    
    We evaluate RAELLA on seven representative DNNs against three state-of-the-art PIM accelerators. Compared to other low-accuracy-loss PIM accelerators, RAELLA improves energy efficiency by up to $4.9\times$ (geomean $3.9\times$) and throughput by up to $3.3\times$ (geomean $2.0\times$). Compared to \WeightCountLimited{} and \SumFidelityLimited{} accelerators that require DNN retraining to recover accuracy, RAELLA provides similar efficiency and throughput while avoiding expensive DNN retraining. 

\mysection{Background and Motivation}
    We first give a brief overview of DNN inference, Processing-In-Memory (PIM), and slicing to lower the resolution of analog operands. We then explore how ADCs limit PIM, how to reduce ADC energy, and the limitations of prior approaches.
    
    \mysubsection{Deep Neural Network (DNN) Inference}
    Modern DNNs are dominated by matrix-vector operations in convolutional and fully connected layers~\cite{efficient_processing_of_dnns}. For inference, 8-bit (8b) per-channel quantized DNNs with 8b inputs/weights and 16b partial sums (psums) are widely available and can achieve high accuracy~\cite{Rokh2022ACS, data_free_quant, Zhao2020LinearSQ, quant_whitepaper, Pruning_Quant_Survey}. RAELLA supports this type of quantization.
    
    DNNs may have billions of multiply-accumulate operations (MACs) over millions of weights~\cite{ResNet}. This makes PIM an attractive choice for DNN inference acceleration. PIM can operate directly in weight memory to reduce data movement~\cite{PRIME} and use Resistive-RAM (ReRAM) for dense and efficient analog compute.

    \mysubsection{ReRAM Properties}
    The functional unit of PIM systems is the ReRAM crossbar. ReRAM crossbars accelerate DNN layers by computing dense in-memory matrix-vector multiplications. Furthermore, ReRAMs are small and offer high storage density~\cite{NVMExplorer}. This density allows ReRAM-based systems to store and run on-chip pipelines that compute DNN layers sequentially~\cite{ISAAC, PipeLayer} without costly accesses to off-chip memory~\cite{efficient_processing_of_dnns}. A disadvantage of ReRAM is high write energy~\cite{reram_survey}. Write cost is amortized in inference as ReRAM is nonvolatile, so written weights can be reused for many inferences~\cite{ISAAC}. 

\begin{figure}
    \includegraphics[width=\linewidth]{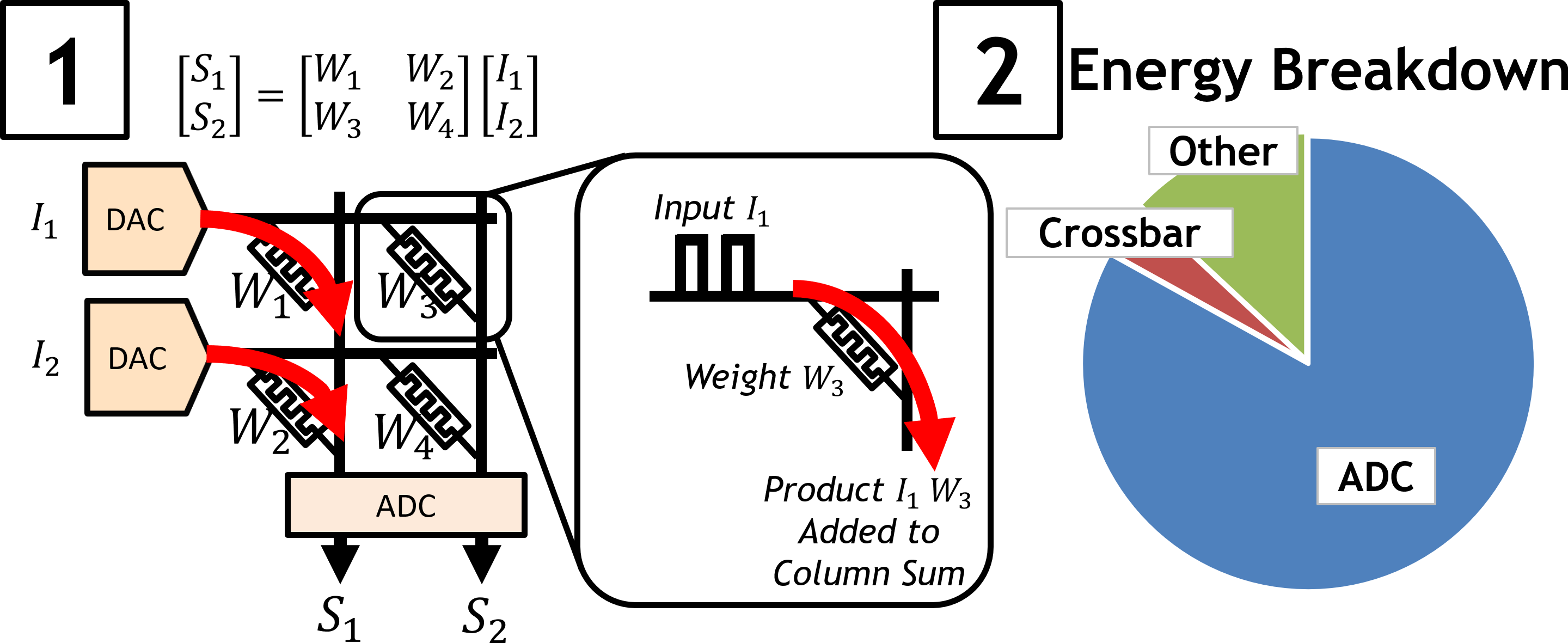}
    \centering
    \caption{Basic PIM crossbar. \square[1] $2\times2$ MVM. Each operand is a single slice. \square[2] Energy breakdown of an ISAAC-based design.}
    \label{fig:xbar}
    \end{figure}

    Fig.~\ref{fig:xbar} shows a basic $2\times2$ matrix-vector multiplication executed on a $2\times2$ ReRAM crossbar. A matrix of weights $W$ is programmed in the ReRAMs. Elements of the input vector $I$ are fed to digital-to-analog converters (DACs), which convert the inputs to analog values. Each ReRAM device multiplies the input on the row with its programmed weight. Products are accumulated in each column to produce analog \emph{\columnsum{}s}, which are converted by an analog-to-digital converter (ADC) to produce the digital result $S$.


    ReRAMs have been shown to be programmable with up to 5b~\cite{High_Precision_Programming_Variation_Intolerant} and 512$\times$512 crossbars have been shown to compute up to 8b \columnsum{}s~\cite{dot_product_engine} under analog noise limitations. These resolution limits necessitate \emph{slicing} to compute higher-resolution DNN layers.

    \mysubsection{Arithmetic with Slices}
    To run 8b DNN inference using lower-resolution devices, PIM architectures partition, or \emph{slice}, input and weight bits into \emph{\inputslice{}s} and \emph{\weightslice{}s}. A slice is a subset of bits from an operand, and multiplying two slices yields a \emph{\slicedproduct{}}.
    
    There are two types of slicing. Temporal slicing processes slices in separate cycles (\eg bit-serial being the extreme with one bit per slice) and spatial slicing processes slices in separate ReRAMs across parallel crossbar columns. In most PIM accelerators that slice, temporal slicing is used for inputs and spatial slicing is used for weights. We refer to a vector of weights and their slices as a \emph{weight filter} if they are mapped to the same set of columns in one crossbar and they contribute to one dot product for a DNN layer.

    Table~\ref{tab:slicing} shows an example of sliced arithmetic. Each \weightslice{} is mapped spatially to one crossbar column, while each \inputslice{} is processed temporally in one cycle. For each column and cycle, an ADC converts the column sum. The result is shifted and added digitally, allowing PIM architectures to calculate full 16b psums despite low-resolution analog limitations~\cite{ISAAC, fundamental_limits_of_crossbar_precision}.

    Table~\ref{tab:slicing} shows tradeoffs relating to slicing. Many costs increase with more slices: each additional \inputslice{} increments $Cycles/Input$ while each additional \weightslice{} increments $Columns/Weight$. $ADC\ Converts$ scales with the product of input and weight slice counts. The benefit of more slices is that we can use fewer bits per slice, thus reducing MAC resolution and required ADC resolution. We can also decrease $ADC\ Converts$ by using larger crossbars that accumulate more analog values across more rows, but this also increases the required ADC resolution.
    
    \begin{table}
        \centering
        \begin{tabular}{@{}lcccccc@{}}
            \multicolumn{6}{c}{\textbf{Dot Product:} 2b input \(i_hi_l\) \(\cdot\) 2b weight \(w_hw_l\)} \\
            \midrule
            \multicolumn{2}{@{\extracolsep{\fill}}l}{\textbf{Sliced Input}} & & \ding{51} & & \ding{51} & \\
            \multicolumn{2}{@{\extracolsep{\fill}}l}{\textbf{Sliced Weight}} & & & \ding{51} & \ding{51} & \\
            \midrule
            Cycle & Column & \\ 
            1 & 1 & \(i_hi_l\cdot w_hw_l\) & \(i_h\cdot w_hw_l\) & \(i_hi_l\cdot w_h\) & \(i_h\cdot w_h\) \\
            1 & 2 & - & - & \(i_hi_l\cdot w_l\) & \(i_h\cdot w_l\) \\
            2 & 1 & - & \(i_l\cdot w_hw_l\) & - & \(i_l\cdot w_h\) \\
            2 & 2 & - & - & - & \(i_l\cdot w_l\) \\
            \midrule
            \multicolumn{2}{@{\extracolsep{\fill}}l}{\textbf{Bits/MAC}} & 4 & 2 & 2 & 1 \\
            \multicolumn{2}{@{\extracolsep{\fill}}l}{\textbf{Converts/MAC}} & 1 & 2 & 2 & 4 \\
            \end{tabular}
        \caption{How Slicing Works \& Tradeoffs. A 2b input/weight are multiplied and each may be sliced into two 1b slices. High and low order bits are \(i_h,w_h\) and \(i_l,w_l\). Each column/cycle computes the \slicedproduct{} shown. More slices reduce bits/slice and bits/MAC, permitting a cheaper, lower-resolution ADC. However, cycles, columns, and ADC converts are needed to process each slice. More slices increase ADC Converts/MAC.
        }
        \vspace{-3mm}
        \label{tab:slicing}
    \end{table}

    \vspace{-1mm}
    \mysubsection{ADCs Limit PIM Accelerators}
    Fig.~\ref{fig:xbar} shows the power breakdown of an 8b PIM architecture based on the foundational ISAAC~\cite{ISAAC}. PIM crossbars are dense and efficient, but are limited by ADC costs. Crossbars can compute 8b MACs with $<$ 100fJ, but overall energy is dominated by ADCs. Crossbars are dense, but architectures can spend 5~\cite{TIMELY} to 50~\cite{ISAAC} times more area on ADC than crossbars. Crossbars can compute with high parallelism, scaling to 1024 rows~\cite{1024_1024_temporal_driver_pulse_train}, but the area and energy of ADCs scale exponentially with resolution~\cite{ADC_Scaling_Murmann}. Prior work has been limited to as few as 16 activated rows~\cite{SRE} to reduce \columnsum{}s and ADC resolution requirements.

    As ReRAM is dense and low power, RAELLA trades off more ReRAM for lower-resolution ADCs. Furthermore, by reducing resolution, we use more crossbar rows/columns with less ADC area/energy scaling. This higher parallelism yields higher throughput and efficiency for the full RAELLA accelerator.

    \mysubsection{Reducing ADC Cost}
    To run efficient PIM inference, we must reduce ADC area and energy. To do so, we present the \IronLaw{} of ADC energy.\footnote{Inspired by the Iron Law~\cite{wiki:Iron_law_of_processor_performance} and titanium-based ReRAM devices~\cite{reram_device_you_use_tiox}.}\footnote{While ADC energy is the focus here, a similar analysis can be performed for area by substituting \emph{Converts/MAC} with \emph{\#ADCs/Throughput}.} Table~\ref{tab:iron} shows the \IronLaw{} equation for ADC energy and breaks down its factors. ADC energy is the product of four terms: 
    \begin{itemize}
    \item \emph{Energy/Convert} is determined by ADC efficiency and scales exponentially with ADC resolution~\cite{ADC_Scaling_Murmann}.\footnote{\emph{Energy/Convert} can also be reduced with clever new ADC designs, but there is an efficiency limit~\cite{ADC_efficiency_max} due to analog noise. This requires innovations on both the ADC and architecture sides.}
    \item \emph{Converts/MAC} is determined by the number of crossbar rows, input slices, and weight slices.
    \item \emph{MACs/DNN} is determined by the DNN workload.
    \item \emph{1/Utilization} corresponds to how many crossbar rows are used by the DNN. A utilization of one means all rows used.
    \end{itemize}

    Given these factors, Table~\ref{tab:iron} shows how to reduce ADC energy by changing hardware attributes. First, notice the tradeoff generated by \emph{Energy/Convert} and \emph{Converts/MAC} in the first/second rows of the table. Although it may seem that slicing and resizing the crossbar can directly reduce ADC energy, this approach has limited benefits. This is because, to reduce \emph{Converts/MAC}, we must either (1) increase the crossbar rows and compute more sliced products per ADC convert, (2) increase bits per \weightslice{}, which reduces the number of columns needed to store each weight and reduces the number of ADC converts needed to process each column, or (3) increase bits per \inputslice{}, which reduces the number of cycles required and ADC converts to process \columnsum{}s over all cycles. The limitation, however, is that in all cases we will accumulate larger and higher-resolution \columnsum{}s. To preserve fidelity, a higher-resolution ADC is needed, which increases \emph{Energy/Convert} and negates our benefits. The converse is true for reducing \emph{Energy/Convert}; preserving high fidelity requires increasing \emph{Converts/MAC}.
    
    The final column of Table~\ref{tab:iron} shows the consequences of reducing each of the \IronLaw{} terms. Of the six consequences, three are ineffective for reducing ADC energy. \emph{Converts/MAC} and \emph{Energy/Convert} trade off with each other. \emph{1/Utilization} cannot be reduced below one.

    \begin{table*}
        \centering
        \vspace*{-2mm}
        \begin{tabular}[b]{lllll} 
            \multicolumn{5}{c}{\noindent\fbox{\textbf{The Titanium Law}:  \(\frac{ADC Energy}{DNN}=\frac{Energy}{Convert} \times \frac{Converts}{MAC} \times \frac{MACs}{DNN} \times \frac{1}{Utilization}\)}} \\
            \multicolumn{5}{c}{}\\ 
            \textbf{Term} & \shortstack[l]{\textbf{Hardware} \\ \textbf{Attribute}} & \textbf{How to Reduce} & \textbf{Tradeoff} & \textbf{Consequence} \\
            \midrule[1.5pt]
            \multirow{2}{*}{\emph{\emph{Energy/Convert}}} & \multirow{2}{*}{\shortstack[l]{ADC \\ Resolution}} & \multirow{2}{*}{\shortstack[l]{Reduce ADC \\ Resolution}} & Fewer Crossbar Rows or Bits/Slice & High \emph{Converts/MAC} \\ \cmidrule{4-5} 
            & & & \circle[S] Fidelity Loss, Psum Errors & \circle[S]\textbf{Accuracy Loss or Retraining} \\
            \midrule
            \multirow{2}{*}{\emph{Converts/MAC}} & \multirow{2}{*}{\shortstack[l]{Crossbar \\ Rows}} & \multirow{2}{*}{\shortstack[l]{Increase Crossbar \\ Rows or Bits/Slice}} & High-Resolution ADC & High \emph{\emph{Energy/Convert}} \\ \cmidrule{4-5} 
            & & & \circle[S] Fidelity Loss, Psum Errors & \circle[S]\textbf{Accuracy Loss or Retraining} \\
            \midrule
            \emph{MACs/DNN} & \# Weights & Prune/Reshape Weights & \circle[W] Eliminated/Changed Weights & \circle[W]\textbf{Accuracy Loss or Retraining} \\
            \midrule
            \emph{1/Utilization} & Mapping & Improve Mapping & Flexibility Cost, Utilization $\le$ 1& Limited Benefits \\
            \midrule
        \end{tabular}
        \caption{\label{tab:iron} The \IronLaw{} of ADC energy and how to reduce ADC energy components. Of the possible consequences, three are ineffective, and three cause accuracy loss or require DNN retraining. \SumFidelityLimited{} architectures choose \circle[S] marked cells and \WeightCountLimited{} architectures choose \circle[W] marked cells.}
        \vspace{-5mm}
    \end{table*}

    \renewcommand{\arraystretch}{0}
    \begin{table}
        \centering
        \begin{tabular}{@{}lllll@{}}
            \shortstack[l]{Architecture \\\ \\\ } & \shortstack[l]{High-Cost\\ ADC} & \shortstack[l]{Limits\\Weight\\Count} & \shortstack[l]{Fidelity\\Loss} & \shortstack[l]{Needs DNN \\ Retraining} \\
            \toprule[1.5pt]
            ISAAC~\cite{ISAAC} & \textcolor{purple}{\textbf{Yes}} & - & - & \textcolor{teal}{\textbf{No}} \\
            \midrule
            AtomLayer~\cite{AtomLayer} & \textcolor{purple}{\textbf{Yes}} & - & - & \textcolor{teal}{\textbf{No}} \\
            \midrule
            FORMS~\cite{FORMS} & \textcolor{teal}{\textbf{No}} & \textcolor{purple}{\textbf{Yes}} & - & \textcolor{purple}{\textbf{Yes}} \\
            \midrule
            SRE~\cite{SRE} & \textcolor{teal}{\textbf{No}} & \textcolor{purple}{\textbf{Yes}} & - & \textcolor{purple}{\textbf{Yes}} \\
            \midrule
            ASBP~\cite{ASBP} & \textcolor{teal}{\textbf{No}} & \textcolor{purple}{\textbf{Yes}} & - & \textcolor{purple}{\textbf{Yes}} \\
            \midrule
            TIMELY~\cite{TIMELY} & \textcolor{teal}{\textbf{No}} & - & \textcolor{purple}{\textbf{High}} & \textcolor{purple}{\textbf{Yes}} \\
            \midrule
            PRIME~\cite{PRIME} & \textcolor{teal}{\textbf{No}} & - & \textcolor{purple}{\textbf{High}} & \textcolor{purple}{\textbf{Yes}} \\
            \midrule
            RAELLA & \textcolor{teal}{\textbf{No}} & - & \textcolor{teal}{\textbf{Low}} & \textcolor{teal}{\textbf{No}} \\
            \midrule
        \end{tabular}
        \caption{Comparison to prior works. Previous approaches pay high ADC costs or use strategies that cause DNN accuracy loss, requiring retraining to recover.}
        \label{tab:table_I}
    \end{table}    

\begin{figure}[]
        \centering
        \includegraphics[width=\linewidth]{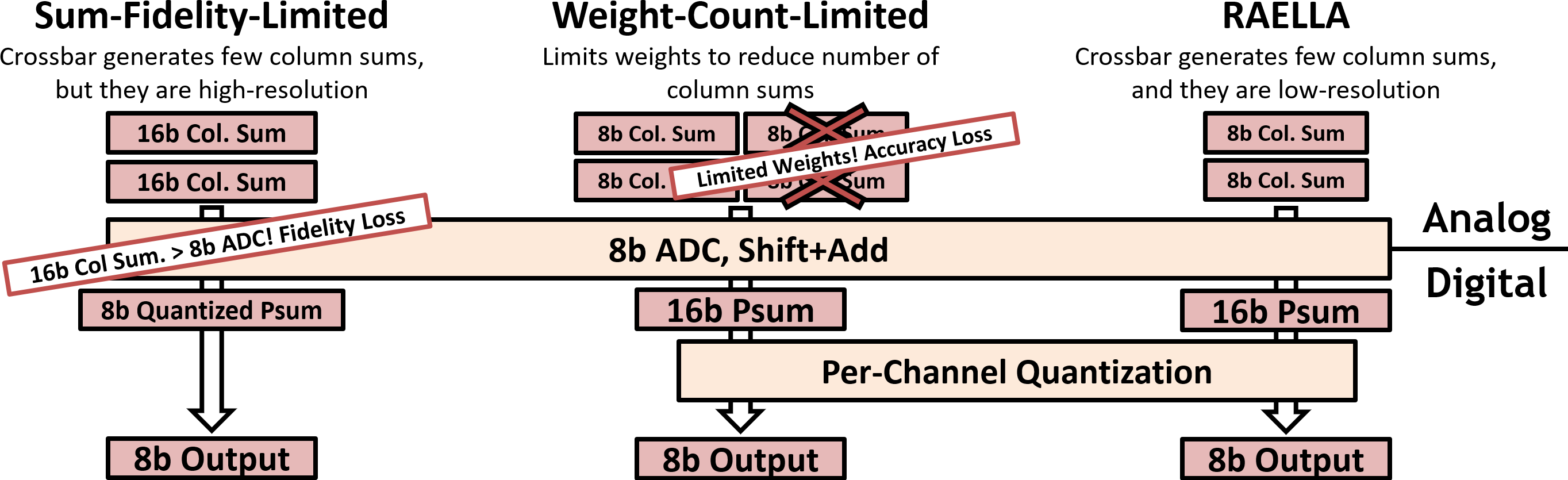}
        \caption{Loss-causing architectures alongside RAELLA. Although they decrease \emph{Converts/MAC}, \SumFidelityLimited{} architectures lose fidelity at the ADCs and force hardware-restricted quantization. \WeightCountLimited{} architectures limit DNN weights. RAELLA's arithmetic and slicing strategies maintain high fidelity with low \emph{Converts/MAC}.}
        \label{fig:why_others_cause_retraining}
    \end{figure}

    Architectures that reduce the ADC energy choose options that end in the consequence \textbf{Accuracy Loss or Retraining}. Fig.~\ref{fig:why_others_cause_retraining} shows how these architectures change DNN operands and lose accuracy. \WeightCountLimited{} architectures, in the \circle[W]-marked cells of Table~\ref{tab:iron}, prune/reshape DNN weights to lower \emph{MACs/DNN}. Unfortunately, changing weights causes DNN accuracy loss unless the DNN is retrained. On the other hand, \SumFidelityLimited{} architectures, in the \circle[S]-marked cells, use more rows, more bits per input/weight slice, and low-resolution ADCs to reduce both \emph{Converts/MAC} and \emph{MACs/DNN}. But because they generate large, high-resolution \columnsum{}s and use low-resolution ADCs, they lose column sum fidelity. This creates errors in outputs and causes accuracy loss unless the DNN is requantized and retrained.
    
    Counterintuitively, 8b ADCs are not always sufficient for 8b-quantized DNNs in \SumFidelityLimited{} architectures. Psums are 16b after MACs of 8b inputs and weights, and high-accuracy linear quantization strategies need the full 16b psum range~\cite{Zhao2020LinearSQ,Wu2020IntegerQF,Rokh2022ACS}, so we would like all 16b to have high fidelity. \SumFidelityLimited{} architectures may generate $>8b$ \columnsum{}s and capture them with an 8b ADC. When the ADCs of these architectures lose bits from \columnsum{}s, they lose bits from the overall psum. This limits high-accuracy quantization strategies to using only the subset of bits that the hardware calculated, rather than the full 16b. This hardware-enforced limitation can cause accuracy loss.

    \mysubsection{Motivation}
    Prior works combat accuracy loss by retraining DNNs. FORMS~\cite{FORMS}, a \WeightCountLimited{} architecture, achieves a $2.0\times$ \emph{MACs/DNN} reduction on ResNet18 by pruning and retraining. TIMELY~\cite{TIMELY}, a \SumFidelityLimited{} architecture, achieves up to a $512\times$ \emph{Converts/MAC} reduction over~\cite{ISAAC} by using large crossbars and many bits per input/weight slice. However, TIMELY also loses 16b of fidelity from each column sum and recovers accuracy with DNN requantization and retraining. Table~\ref{tab:table_I} shows a gap in recent PIM works: some PIM architectures are inefficient and do not reduce high ADC costs, while others that reduce ADC costs cause DNN accuracy loss and retrain to compensate.
    
    Retraining DNNs can be a challenge due to high computational cost~\cite{DNN_energy}, cumbersome hyperparameter tuning~\cite{hyperparameter_tuning}, and the potential lack of access to training datasets~\cite{deepface_proprietary, dalle_proprietary}. Additionally, highly efficient DNNs such as highly-reduced-precision models~\cite{intel_low_precision_dnn, low_precision_rnn, ibm_low_precision_inference} often depend on their own training/quantization procedures. If an architecture requires different training/quantization procedures, it may be difficult or impossible to run these cutting-edge DNNs. The motivation behind RAELLA is to deliver efficient inference and avoid accuracy loss without retraining or modifying DNNs.

    \mysection{RAELLA: Low Resolution, High Fidelity}

    To be efficient, we would like to reduce ADC resolution. But if column sum resolution is greater than ADC resolution, we lose fidelity and DNN accuracy. We identify three architectural tradeoffs that create high-resolution \columnsum{}s:
    
    \begin{itemize}
        \item More sliced products per ADC convert $\xrightarrow[]{}$ fewer ADC converts, higher-resolution \columnsum{}s.
        \item More bits per weight slice $\xrightarrow[]{}$ fewer weight columns, fewer ADC converts, higher-resolution \columnsum{}s.
        \item More bits per input slice $\xrightarrow[]{}$ fewer input cycles, fewer ADC converts, higher-resolution \columnsum{}s.
    \end{itemize}

\begin{figure}[]
        \centering
        \includegraphics[width=\linewidth]{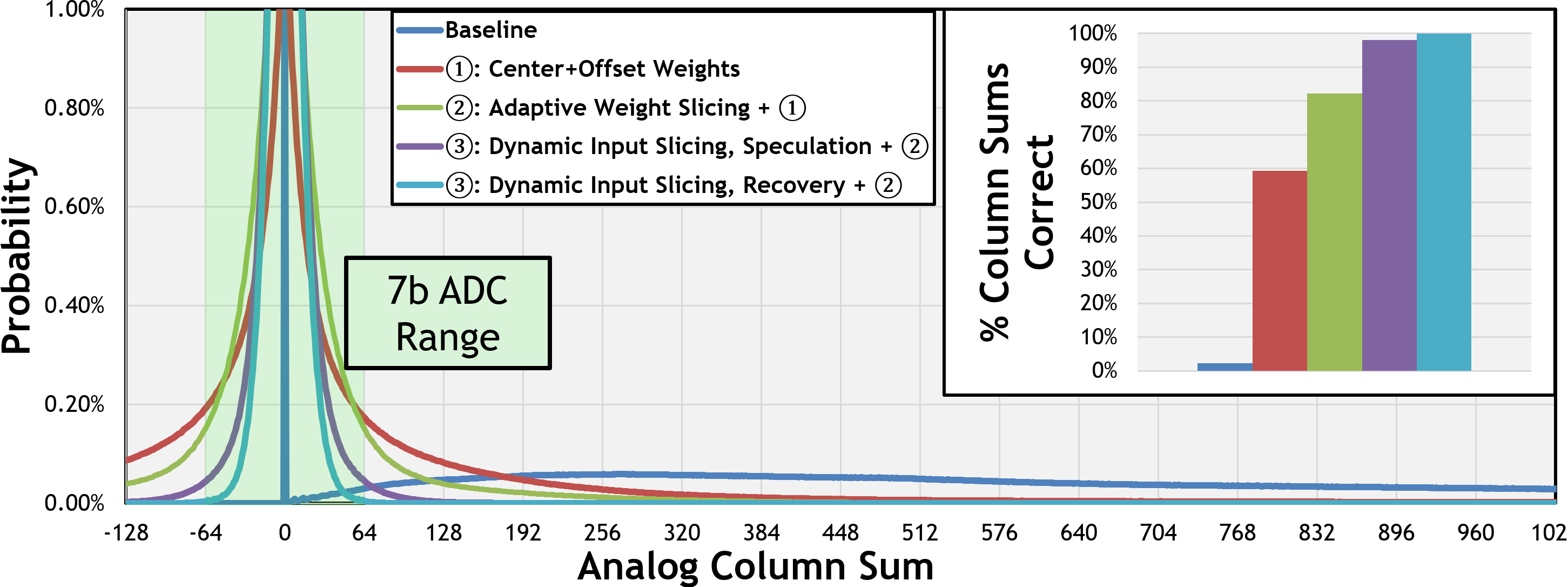}
        \caption{Column sum distribution with each of RAELLA's strategies while running ResNet18 on ImageNet. RAELLA reduces column sum resolution from 17b to 7b and reduces ADC saturation rate from 98\% to 0.1\%.}
        \label{fig:colsum_pdf}
    \end{figure}

    Here, we give an overview of RAELLA's strategies targeting these three tradeoffs. We start with a baseline that uses a $512\times512$ crossbar and 4b input/weight slices. Shown in Fig.~\ref{fig:colsum_pdf}, this setup will produce a very wide distribution of column sums that range from zero to tens of thousands. It requires 17b to represent these column sums. RAELLA's strategies tighten the column sum distribution until it can be represented with a signed 7b range of $[-64,64)$.

    To capture this 7b range without losing fidelity, we set RAELLA's ADC to always capture the seven least-significant bits (LSBs) of column sums. That is, if a single crossbar row is on and produces a \slicedproduct{} of one, the ADC will read the column sum and output the value one. This small step size preserves full fidelity for in-range column sums.\footnote{This strategy contrasts with the approach of many \SumFidelityLimited{} architectures, which drop LSBs from computations~\cite{TIMELY, PRIME, CASCADE} While dropping LSBs permits a lower saturation chance, it also necessarily loses fidelity in every psum.} The drawback is that a small step size means a small range; the ADC saturates and loses fidelity if the column sum is outside $[-64,64)$. With 4b input/weight slices, even a single crossbar row can produce \slicedproduct{}s up to $(2^4-1)^2=225$, which would be saturated at 63. RAELLA must avoid saturation while summing 512 rows at once.
    
    By reshaping the \columnsum{} distribution, RAELLA's strategies reduce the probability of saturation to near-zero. This is how RAELLA achieves high fidelity with a low-resolution ADC: RAELLA's ADC only loses fidelity if column sums are large, and the following strategies make column sums small. For each strategy, we report the ADC saturation rate running ResNet18 on ImageNet.

    \subsection{Center+Offset Weights}
    
    \mysubsubsection{Problem} Standard ReRAM crossbars compute unsigned \slicedproduct{}s. If each \slicedproduct{} is $\ge0$, then accumulating many \slicedproduct{}s will generate large-valued, high-resolution \columnsum{}s.
    
    \mysubsubsection{Solution} We shift weights by a center value such that approximately half of the weights are above the center and half are below. As a result, when we slice weights and compute with them in a crossbar, approximately 50/50\% of the \slicedproduct{}s come from positive/negative weights. We then sum the signed \slicedproduct{}s in-crossbar. Positive and negative \slicedproduct{}s negate, yielding small-valued \columnsum{}s even as many \slicedproduct{}s are accumulated. To maximize the beneficial negation that occurs, Center+Offset chooses centers that balance the magnitude of positive/negative slices in each crossbar column.
    
    \mysubsubsection{Tradeoff} RAELLA trades off higher crossbar area to implement signed arithmetic in-crossbar. Crossbars are dense, so the area tradeoff is worthwhile to reduce ADC cost. RAELLA also uses additional storage and low-cost digital circuitry to store and process center values.
    
    \mysubsubsection{Result} With Center+Offset weights, the \columnsum{} distribution labeled \circle[1] in Fig.~\ref{fig:colsum_pdf} is signed and centered around zero. Column sum resolution is $\le7$b 59.2\% of the time.

    \subsection{Adaptive Weight Slicing}
    
    \mysubsubsection{Problem} More bits per weight slice increase the values stored in weight slices, raising column sum values and resolutions.
    
    \mysubsubsection{Solution} Shown in Fig.~\ref{fig:mp_description}, RAELLA adaptively slices weights at compilation time. We can reduce the average values stored in weight slices and reduce column sum resolution by using fewer bits in each weight slice. However, additional weight slices increase the storage footprint and number of ADC converts by increasing the number of columns, so we would like to minimize the number of slices used. During compilation, we measure errors caused by fidelity loss. We choose the number of bits in each weight slice to control errors and minimize the number of slices used. RAELLA can use a different number of bits for each slice, but all weights in a layer use the same slicing.
    
    \mysubsubsection{Tradeoff} RAELLA trades off storage density, ADC converts, and compilation-time preprocessing. ReRAMs and ADC converts needed increase with number of weight slices. RAELLA uses a simple preprocessing strategy and reuses DNN weights for many inferences to minimize preprocessing costs.

    \mysubsubsection{Result} With Adaptive Weight Slicing, the \columnsum{}s labeled \circle[2] in Fig.~\ref{fig:colsum_pdf} are more tightly distributed. Column sum resolution is $\le7$b 82.1\% of the time.

    \subsection{Dynamic Input Slicing}
    
    \mysubsubsection{Problem} More bits per input slice increase the values of input slices, raising column sum values and resolutions.

    \mysubsubsection{Solution} Shown in Fig.~\ref{fig:speculation}, RAELLA dynamically slices the inputs at runtime. RAELLA can use fewer bits per input slice to reduce column sums. However, this requires more cycles and more ADC converts. RAELLA uses a dynamic strategy by speculating with an efficient approach of more bits per \inputslice{}. RAELLA recovers from large-column-sum saturation errors by using fewer bits per \inputslice{}. This approach achieves high efficiency from speculation and high fidelity from recovery.

    \mysubsubsection{Tradeoff} RAELLA trades off throughput and crossbar energy. While typically speculation is used to increase speed, RAELLA's speculation trades off speed to gain efficiency. Extra cycles are needed to run both speculation and recovery. Additionally, RAELLA's crossbars consume energy for both speculation and recovery. As crossbars are high-throughput and efficient, it is worth the cost to reduce the ADC overhead.
    
    \mysubsubsection{Result} With Dynamic Input Slicing, speculation and recovery \columnsum{} distributions labeled \circle[3] in Fig.~\ref{fig:colsum_pdf} are further tightened. In speculation and recovery cycles, \columnsum{} resolution is $\le7$b 98.0\% and 99.9\% of the time, respectively.

    \subsection{Accepting Fidelity Loss}

    \mysubsubsection{Problem} With all of RAELLA's optimizations, the \columnsum{} resolution can still be greater than ADC resolution. We use a 7b ADC and produce $>$7b \columnsum{}s 0.1\% of the time. These cause the ADC output to saturate at its min/max of -64/63 and propagate incorrect values to the psum.

    \mysubsubsection{Solution} DNNs are inherently noise-tolerant~\cite{noise_tolerance, quantization_error_is_gaussian_noise} so a low error rate is acceptable. Table~\ref{tab:Accuracy Comparison} shows that RAELLA's fidelity errors cause low loss for a variety of DNNs.

    RAELLA uses a 512-row crossbar and a 7b ADC. Even with minimal 1b input and 1b weight slices, column sum resolution may be 9b, so it is impossible to guarantee perfect fidelity. With minimal weight slice sizing, RAELLA reduces ADC-related fidelity errors to a rate on the order of one error in ten million psums, or one incorrect psum per ResNet50~\cite{ResNet} inference.

\begin{figure}
    \includegraphics[width=\linewidth]{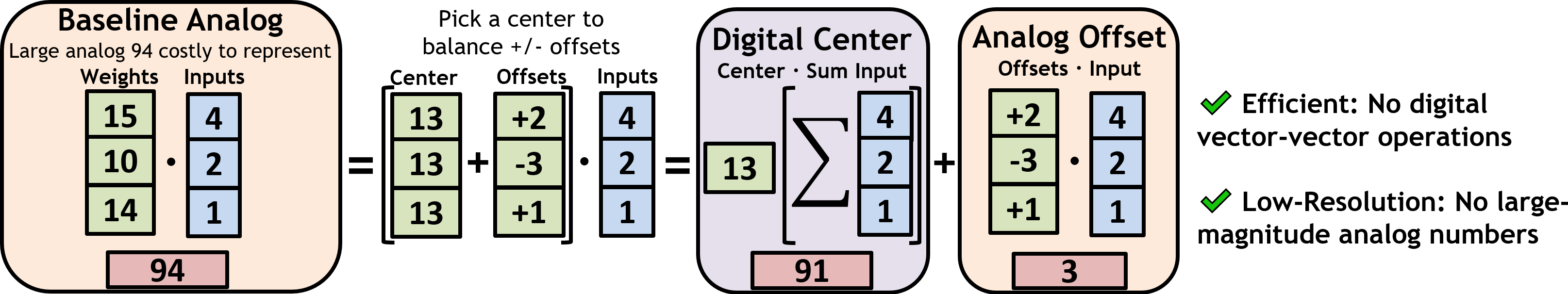}
    \centering
    \caption{Center+Offset weights. Standard dot products create high-resolution values which are difficult to represent in analog. Center+Offset digitally subtracts a center from weights, computing with near-zero-average offsets in-crossbar.}
    \label{fig:median_split}
    \end{figure}

    \mysection{Implementing RAELLA's Strategies}
    \mysubsection{Implementing Center+Offset Weights} \label{2T2R} \label{center_offset}
    Shown in Fig.~\ref{fig:median_split}, we represent DNN weights as a center value plus or minus a small offset. We select centers to make positive and negative weight slices approximately the same magnitude for each column. RAELLA computes signed analog arithmetic, and \slicedproduct{}s from the magnitude-balanced positive/negative weight slices negate to produce near-zero column sums. This allows RAELLA to keep small column sums while accumulating \slicedproduct{}s from many crossbar rows. Meanwhile, RAELLA efficiently processes high-resolution centers in the digital domain. 

    We first discuss why Center+Offset is important for balancing positive/negative weight slices, then show how RAELLA computes arithmetic with Center+Offset encoding and describe how we calculate optimal center values. Finally, we show the hardware for computing dot products with Center+Offset encoded weights. 

    \mysubsubsection{Why Balance Slices}
    While DNN weight distributions are commonly cited as zero-mean~\cite{fidelity_encoding_exploration, fundamental_limits_of_crossbar_precision}, a zero mean for all weight values over a DNN does not necessarily mean any given weight filter is zero-mean, nor that \columnsum{}s are zero-mean. For that, we need each individual crossbar column to have weight slices with a zero mean. This is often not the case, as individual weight filters and columns of slices randomly converge to different distributions.\footnote{When we say “filter” we mean a set of weights from one dot product that fit in one crossbar. An output channel of a DNN layer (or “filter” in the traditional sense) may be partitioned over multiple crossbars if its weights do not fit in a crossbar. The important aspect is that each crossbar column produces a unique column sum distribution, regardless of the characteristics of the overall DNN. To account for this, Center+Offset attempts to balance positive/negative weight slices in each column.}

    A growing body of works are exploring differential encoding, which, like Center+Offset, computes signed analog arithmetic~\cite{2T2R, 2T2R_2, 2T2R_3,  1T2R_Aeris, fundamental_limits_of_crossbar_precision, 1024_1024_temporal_driver_pulse_train, fidelity_encoding_exploration}. Differential encoding uses positive slices to represent positive weights and negative slices to represent negative weights; it can benefit from Center+Offset to balance positive/negative weight slices and reduce column sum resolution.
    
\begin{figure*}
    \includegraphics[width=\linewidth]{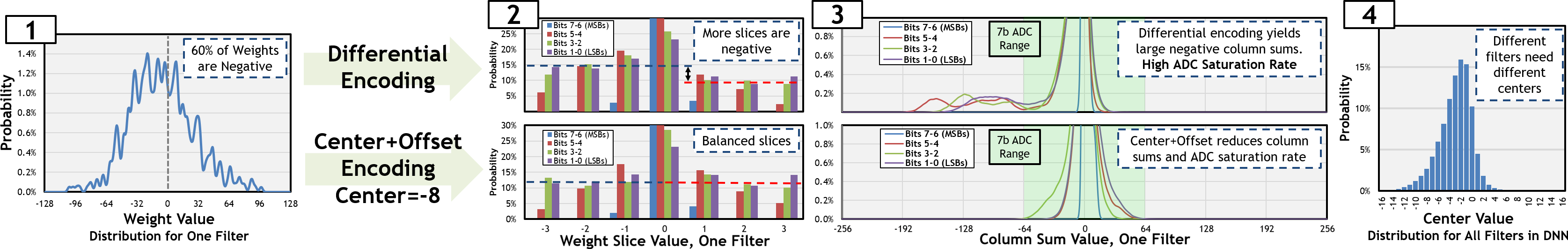}
    \centering
    \caption{Differential vs. Center+Offset Encoding. Distributions for an InceptionV3~\cite{InceptionV3} filter with negative-average weight slices is shown for illustrative purposes. 8b weights / inputs are sliced into four 2b / eight 1b slices. \square[1] Most of the weights in a filter are negative. \square[2] Differential encoding represents negative weights with negative slices, yielding mostly-negative slices for the filter. Center+Offset balances positive/negative slices. \square[3] Dot products with mostly-negative slices yield large negative column sums that cause ADC saturation. Center+Offset reduces column sums. \square[4] Each DNN filter needs a different center.} 
    \label{fig:CO_versus_differentail}
    \end{figure*}

    Center+Offset can be especially beneficial for filters where weight slice distributions have noticeable nonzero averages. This can occur in filters where there is a greater number of negative than positive weights, such as the filter shown in Fig.~\ref{fig:CO_versus_differentail}. Differential encodings represent these mostly-negative weights with mostly-negative slices, yielding a negative average for the slices in each column. After dot products with hundreds of slices, even slight negative averages can accumulate to create large negative \columnsum{}s. This effect can significantly increase ADC saturation and cause DNN accuracy loss, as shown in Table~\ref{tab:Accuracy Comparison}. By balancing positive/negative slices, Center+Offset reduces per-column biases and protects from accuracy loss.

    \mysubsubsection{Center+Offset Arithmetic} Given a weight $w$ and center $\phi$, we calculate positive offset ${w_+=max(w - \phi, 0)}$ and negative offset ${w_-=max(\phi - w, 0)}$. For weights above the center, $w_+$ is the difference between the weight and the center while $w_-$ is zero. The converse is true for weights below the center. Given weight filter $W$ programmed as positive/negative offset vectors $W_+$, $W_-$, RAELLA computes a dot product with input vector $I$ as:
    \begin{equation} \label{eq:offs_prod}
    W \cdot I=\left(\phi\sum I\right) + (W_+ - W_-)I
    \end{equation}

    RAELLA computes Eq.~\ref{eq:offs_prod} at runtime, with $\phi\sum I$ computed digitally and $(W_+ - W_-)I$ in analog.
    


    \mysubsubsection{Calculating Optimal Centers}
    We calculate centers/offsets with one-time preprocessing before programming RAELLA. We calculate a center for each weight filter independently, as weight distributions and optimal centers vary for different weight filters.

    We define an optimization problem to solve for the center value $\phi$. First, we define a slice $S$ as a sequence of inclusive bit indices $[h \dots l]$ from the most to least significant index~$h$ to~$l$ (\eg slice $[7 \dots 4]$ contains the four most significant bits). Then, we define a slicing function $D(h, l, x)$ that crops signed number $x$ to contain the bits from indices $h$ to $l$ (shifted so bit $l$ is the least significant position), preserving the sign. Given a weight filter $W$ and slices $S_{i \in \{1,2,\dots,N\}} = [h_i \dots l_i]$, we solve for the center $\phi$ of $W$ as follows:
    \begin{equation}\label{eqn:center}
        \argmin_{\phi \in \{1, 2, \dots, 255\}} \sum_{i=1}^{N}2^{l_i}\left( \sum_{w \in W}D(h_i, l_i, w-\phi)\right)^4,
    \end{equation}
    
    where $N$ is the total number of slices, and $w$ is a weight in $W$. Eq.~\eqref{eqn:center} balances positive/negative values in each column of \weightslice{}s, assigning higher costs for columns with larger nonzero sums. The inner sum \(\left( \sum_{w \in W}D(h_i, l_i, w-\phi)\right)^4\) calculates the cost for a single column, equal to the sum of \weightslice{}s in the column raised to the power of four. Four is chosen empirically; we find that too low a power does not sufficiently penalize large sums, while too high a power overvalues the largest-sum column and fails to consider all columns. The outer sum \(\sum_{i=1}^{N}2^{l_i}(...)\) weights cost by bit position (\eg the most significant bit in an 8b number has a magnitude of $2^7$ and the cost of a 1b slice containing only this bit would also be scaled by $2^7$) and sums costs for all columns. Costs are weighted by bit position as saturations in higher-order slices have a greater impact on the psum.

    
    
    We calculate centers for each weight filter (\ie a single dot product) in the crossbar independently. A coarser granularity, such as a single center for a full crossbar (\ie 100+ different filters), would not be as effective, as different DNN filters can have different weight distributions and require different centers. 
    
    RAELLA's per-filter centers have the drawback that each center balances multiple columns for one filter, and therefore may not be optimal for any one column. Ideally, per-column centers would be able to precisely zero the average weight slice value for each column. However, this approach is limited by integer precision centers. Consider the case where a column of slices has an average value of $0.4$. We could shift all weight slices in the column by $-1$, but this would worsen the average by shifting it to $-0.6$. Instead, we shift full-precision weight values before slicing, which can reshape (rather than shift) the value distribution for each individual slice. This distribution reshaping can make smaller adjustments to the average weight slice value in each column.

    \begin{figure}
    \includegraphics[width=\linewidth]{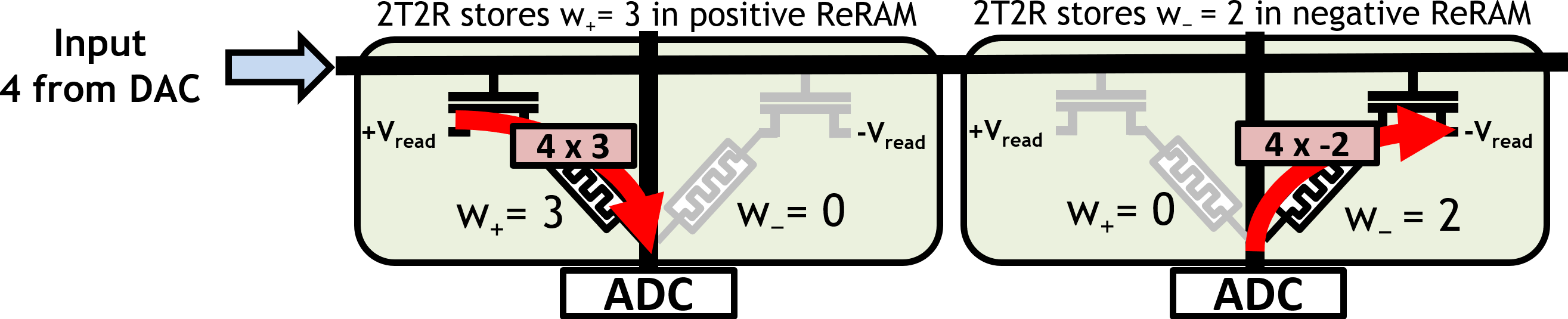}
    \centering
    \caption{2T2R devices compute signed arithmetic in-crossbar. Red shows the magnitude and direction of the current flow. Grayed-out devices are off (set to the high-resistance state).}
    \label{fig:2T2R}
    \end{figure}
    
    \mysubsubsection{Center+Offset In Hardware}
    Computing psums with Eq.~\eqref{eq:offs_prod} requires two terms. The first term, the input sum $\phi\sum{I}$, is calculated digitally. Crossbar columns share input vectors, so the sum calculation is amortized across columns.

    The second term is the vector-vector multiplication with offsets. $(W_+ - W_-)I$ is calculated in analog by the crossbar. RAELLA uses 2T2R devices, shown in Fig.~\ref{fig:2T2R}, to realize analog subtraction in-crossbar.\footnote{Analog subtraction can also be done with circuits~\cite{sram_analog_sub} 1T2R~\cite{1T2R_Aeris}, and SRAMs~\cite{ternary_sram, ternary_sram_2}.}  2T2R, with two ReRAMs (2R) per weight accessed via two access transistors (2T), have been explored as a method to represent signed weights~\cite{2T2R, 2T2R_2, 2T2R_3, stacked_MP, stacked_MP_2}. One ReRAM device is connected to a positive source and the other a negative source, letting 2T2Rs add to or subtract from column sums. For each weight, we program positive/negative offsets $w_+$/$w_-$ into the two ReRAMs. As one offset is zero for any given weight, one ReRAM device is used in each pair. Added ReRAMs and access transistors increase RAELLA's crossbar size, but crossbars are small, and the increase in system area is only $\sim10\%$.

\begin{figure}
    \includegraphics[width=\linewidth]{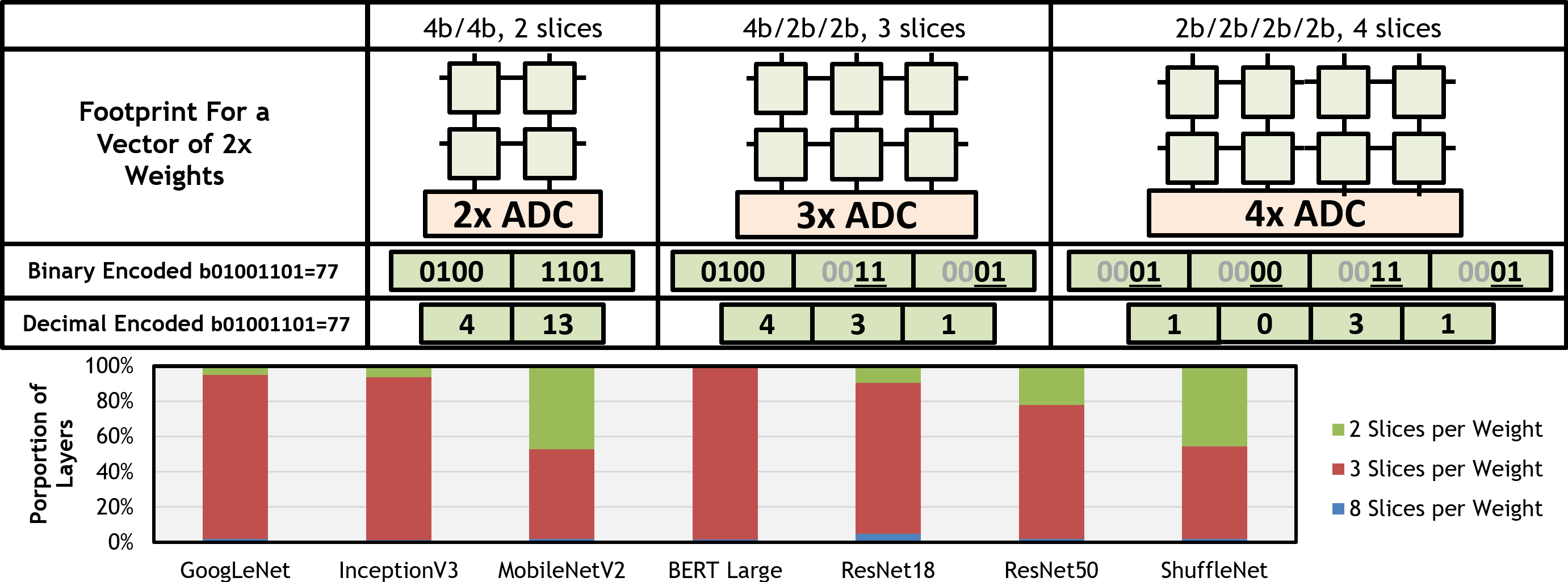}
    \centering
    \caption{\textbf{(Top)} Weight Slice Crossbar Footprints. \textbf{(Bottom)} DNN Per-Layer Weight Slicings. Increasing slice count lowers column sums and saturation chance, but increases \emph{Converts/MAC}. Most layers use three slices per weight.}
    \label{fig:mp_description}
    \end{figure}

    \mysubsection{Implementing Adaptive Weight Slicing} \label{adaptive_weight_slicing}
    Adaptive Weight Slicing minimizes the weight slices used for each DNN layer. It uses as many bits as possible in each slice without excess fidelity loss. Fig.~\ref{fig:mp_description} shows various slicings available to RAELLA. More bits per slice means fewer slices per weight, denser storage, and fewer ADC converts, but more bits also increase the values stored in each weight slice and raise the chance of high-resolution column sums. RAELLA can use a different number of bits for each slice, but all weights in a layer use the same slicing.

\begin{figure}
    \includegraphics[width=\linewidth]{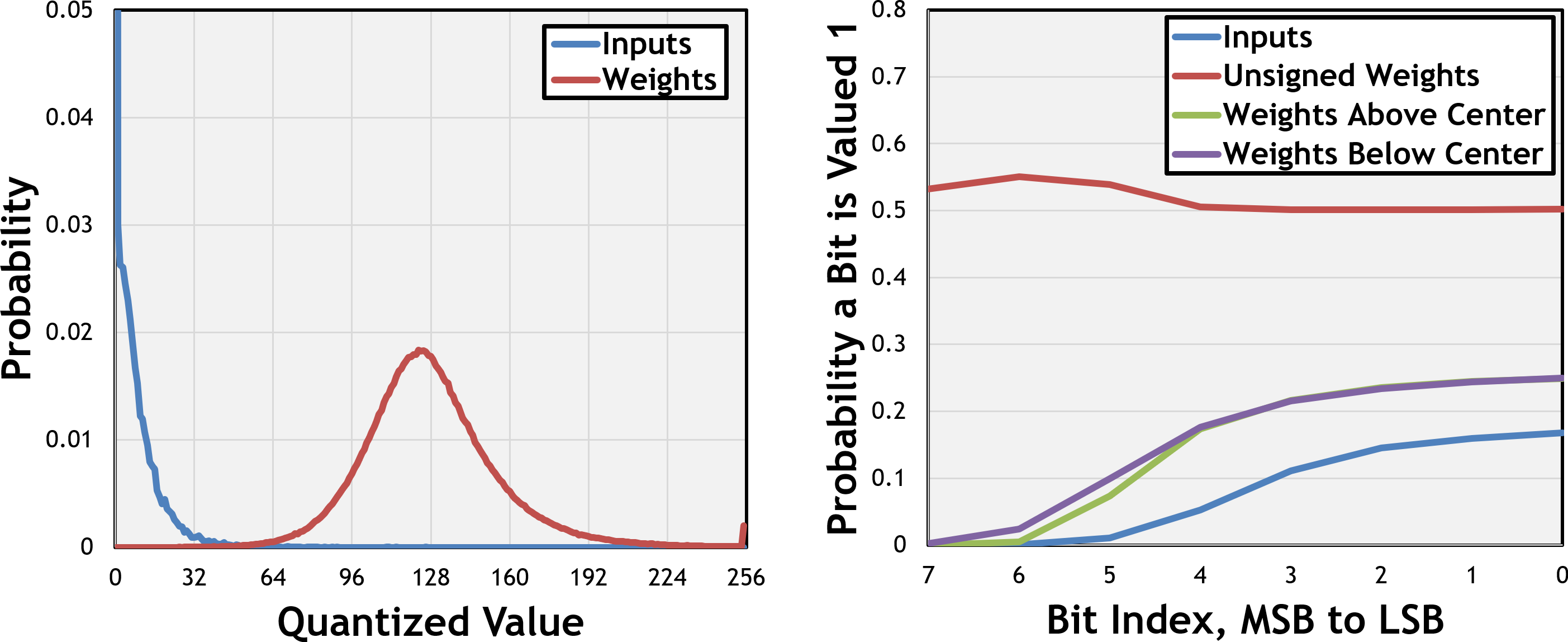}
    \centering
    \caption{\numberlabel[Left] DNN input/weight value distributions without slicing and \numberlabel[Right] per-bit densities. The second-to-last layer of ResNet50 is shown, representing a typical DNN layer. Bell-curve-distributed weights can be split about a center into two similar distributions with sparse high-order bits. Unsigned inputs have naturally-sparse high-order bits. }
    \label{fig:bit_distributions}
    \end{figure}

    The bit density, or probability that a given bit is 1, affects the values in weight slices. Fig.~\ref{fig:bit_distributions} shows per-bit densities for DNN inputs and weights in a typical DNN layer. Input values generally follow right-skewed distributions, yielding sparse high-order bits. Weight values generally follow rough bell curves. When represented with Center+Offset encoding, this also yields sparse high-order bits. Due to sparsity in the high-order weight bits, in most layers, 4b \weightslice{}s can store the highest-order 4b of weights with low values and low \columnsum{}s. Low-order weight bits are denser and usually require a lower 2b per \weightslice{}. Fig.~\ref{fig:mp_description} shows the per-layer slicings of DNNs. Most layers use the 4b-2b-2b setup with three \weightslice{}s: one \weightslice{} for the highest-order four bits and two \weightslice{}s storing two low-order bits each.

    \mysubsubsection{Error Budgets}
    We could choose weight slices to minimize saturation, but this is too conservative. DNNs can tolerate some error, so we would like to allow a small amount of saturation. Unfortunately, it is difficult to predict how slicing impacts saturation and how saturation affects error in DNN layer outputs. Too-large column sums cause saturation, and column sums are affected by input/weight distributions, input/weight slice distributions, correlations between distributions, the number of weights per filter, and random variance. Furthermore, 16b psums are digitally quantized into 8b outputs~\cite{Zhao2020LinearSQ}, which may magnify or shrink the error.

    To capture all these factors, we take an empirical approach. We define the \emph{error budget} as the average magnitude error allowed for nonzero outputs of a layer after outputs are fully computed and quantized to 8b. Only nonzero outputs are counted when calculating average magnitude error to give a more consistent calculation for layers with varying output sparsity.\footnote{This is important when ReLU is folded into quantization. If ReLU zeros an output, it will likely zero any error associated with that output as well. If ReLU zeros many outputs, then average error is lowered while error per nonzero output remains consistent.}

    To calculate error, we use ten inputs chosen randomly from the validation set. It suffices to use so few inputs because changing slicings may change the error by an order of magnitude or more, and these differences are easily detected.\footnote{In fact, this algorithm usually picks the same slicings when testing just one input. If we test with Gaussian noise as input, then slicings match for $>90\%$ of layers.} The order-of-magnitude differences stem from the shape of the column sum distribution (Fig.~\ref{fig:colsum_pdf}). The distribution tails shrink exponentially, so changes in the distribution width (\ie due to slicings) have an exponential effect on the saturation rate.
    
    The error budget is set to 0.09 in our tests, corresponding to one in eleven 8b outputs being off by one on average. After quantization, the errors created by ADC saturation are generally small and cause a low accuracy loss, shown in Table~\ref{tab:Accuracy Comparison}.
    
    \mysubsubsection{Choosing Weight Slices} \label{Choosing weight slices}
    Weight slices are calculated with the preprocessing procedure shown in Algorithm~\ref{alg:programming_algorithm}. Preprocessing occurs once when compiling a DNN for RAELLA, taking 10-1000ms per layer on an Nvidia RTX 2060 GPU. After preprocessing, sliced+encoded weights are programmed to crossbars for use with any number of inferences.
    
    For an M-bit weight and up to N bits per ReRAM, we define a \emph{slicing} as a tuple of integers \(1\leq s_0..s_j\leq N\) such that \(\sum{s_i}=M\). For 8b weights, \(\leq4\)b per ReRAM, slicings include (4b,4b), (2b,1b,1b,4b), and (1b,2b,2b,3b). There are 108 slicings in total. 

    To find the best slicing for a DNN layer, we iterate through all 108 slicings. For each, we Center+Offset encode weights following Section~\ref{2T2R}, simulate the crossbar with ten test inputs, and record error. We choose the slicing that uses the fewest slices and has below-budget error. For slicings with the same number of \weightslice{}s, the lower-error slicing is chosen. 
    
    We use 1b \inputslice{}s when comparing weight slicings. We always use the most conservative 1b per \weightslice{} for the last layer. The last layer has an outsized effect on DNN accuracy~\cite{ibm_low_precision_inference} and a less efficient last-layer slicing has little effect on overall energy/throughput as intermediate layers dominate DNNs (Fig.~\ref{fig:mp_description}).

    \begin{algorithm}[ht]
      \small
      \caption{\hbox{\textbf{Preprocessing Weight Slicing and Centers} \label{alg:programming_algorithm}}}
      \SetAlgoLined
      \DontPrintSemicolon
      \SetKwProg{Fn}{Func}{}{}
      \SetKwFunction{CenterOffsetEncode}{CenterOffsetEncode}
      \SetKwFunction{SlicingLoss}{SlicingLoss}
      \SetKwFunction{RunLayer}{RunLayer}
      \SetKwFunction{GetAllPossibleSlicings}{GetAllPossibleSlicings}
      \SetKwFunction{Mean}{Mean}
      \SetKwFunction{CountSlices}{CountSlices}
      \SetKwFunction{AdaptiveWeightSlice}{AdaptiveWeightSlice}
      \SetKwFunction{OptimizeCenters}{OptimizeCenters}
      \SetKwFunction{Slice}{Slice}
      \SetKwFunction{EncodeDNN}{EncodeDNN}
      \SetKwFunction{IdealRunLayer}{IdealRunLayer}
      \SetKwFunction{SolveOptimizationProblem}{SolveOptimizationProblem}
      \SetKwFunction{FindOptimalCenters}{FindOptimalCenters}

      \SetKwFunction{SliceEncodeWeights}{SliceEncodeWeights}
      \SetKwFunction{SimulateCrossbar}{SimulateCrossbar}
      \SetKwFunction{SimulateAnalog}{SimulateAnalog}
      \SetKwFunction{FindBestSlicing}{FindBestSlicing}
      \SetKwFunction{Run}{Run}
      \Fn{\SliceEncodeWeights{layer, testInputs, errorBudget}}{
      \tcc{DNN layer preprocessing. Requires a layer (shape, quantization, weights), test inputs (activations from ten images/tokens in this paper), and a scalar error budget (0.09 in this paper).}
      \hbox{slicing = \FindBestSlicing{layer, testInputs, errorBudget}\;}
      centers = \FindOptimalCenters{layer, slicing}\;
      \KwRet slicing, centers\;
      }\;
      
      \Fn{\FindBestSlicing{layer, testInputs, errorBudget}}{
      \tcc{Implementation of Adaptive Weight Slicing from Sec.~\ref{adaptive_weight_slicing}. 10-1000ms per layer.}
      expectedOutputs = layer.\Run{testInputs}\;
      possibleSlicings = \GetAllPossibleSlicings{}\;
      bestSlicing = possibleSlicings[0]\;
      bestNSlices = \CountSlices{bestSlicing}\;
      bestError = $\infty$\;
        \For{\(slicing \in possibleSlicings\)} {
            \upshape
              centers = \FindOptimalCenters{layer, slicing}\;
              outputs = layer.\SimulateCrossbar{testInputs, slicing, centers}\;
              errors = |expectedOutputs - outputs|\;
              meanError = \Mean{errors[expectedOutputs != 0]}\;
              nSlices = \CountSlices{slicing}\;
              betterSlicing = nSlices < bestNSlices || (nSlices==bestNSlices \&\& meanError < bestError)\;
              \uIf{meanError < errorBudget \&\& betterSlicing} {
                    bestSlicing = slicing\;
                    bestNSlices = nSlices\;
                    bestError = meanError\;
              }
        }
    \KwRet bestSlicing\;
    }\;

      \Fn{\FindOptimalCenters{layer, slicing}}{
      \tcc{Solve Center+Offset Eq.~\eqref{eqn:center}. <1ms per layer. Returns a center for each weight filter.}
      centers = \SolveOptimizationProblem{layer, slicing} \;
      \KwRet centers\;
      }
    \end{algorithm}

    \mysubsubsection{Adaptive Weight Slicing in Hardware}
    Given 4b ReRAMs, each can be programmed with $2^{4}-1$ analog levels. To program 3b or 2b slices, we use the lowest $2^{3}-1$ or $2^{2}-1$ levels. Given a 3b \weightslice{} XXX, this corresponds to programming a device with 0XXX. This is only a restriction of the available range and therefore does not require a change to ReRAMs. Crossbars already need shift+add circuits to add column sums across weight and input slices; adaptive slicing requires only changing the shift+add pattern.
    
    The main overhead depends on the number of weight slices. Each additional weight slice increases required ReRAMs and ADC converts. RAELLA can use between two weight slices (4b/slice, most efficient) and eight weight slices (1b/slice, least efficient). Most layers use three weight slices.
    
    \mysubsection{Implementing Dynamic Input Slicing}~\label{Speculation}
    Dynamic Input Slicing balances high-efficiency more-bit \inputslice{}s and high-fidelity fewer-bit \inputslice{}s. We would like to minimize the input slices and thus ADC converts, while also avoiding fidelity loss due to high-resolution column sums. Unlike with weights, the input slicing can be changed at runtime. This allows us to speculate with an efficient, aggressive slicing and recover with a conservative slicing. In speculation, RAELLA uses three \inputslice{}s, which has high efficiency but a higher chance of creating large, high-resolution \columnsum{}s. In recovery, RAELLA uses the most conservative eight 1b input slices.

\begin{figure}
    \includegraphics[width=\linewidth]{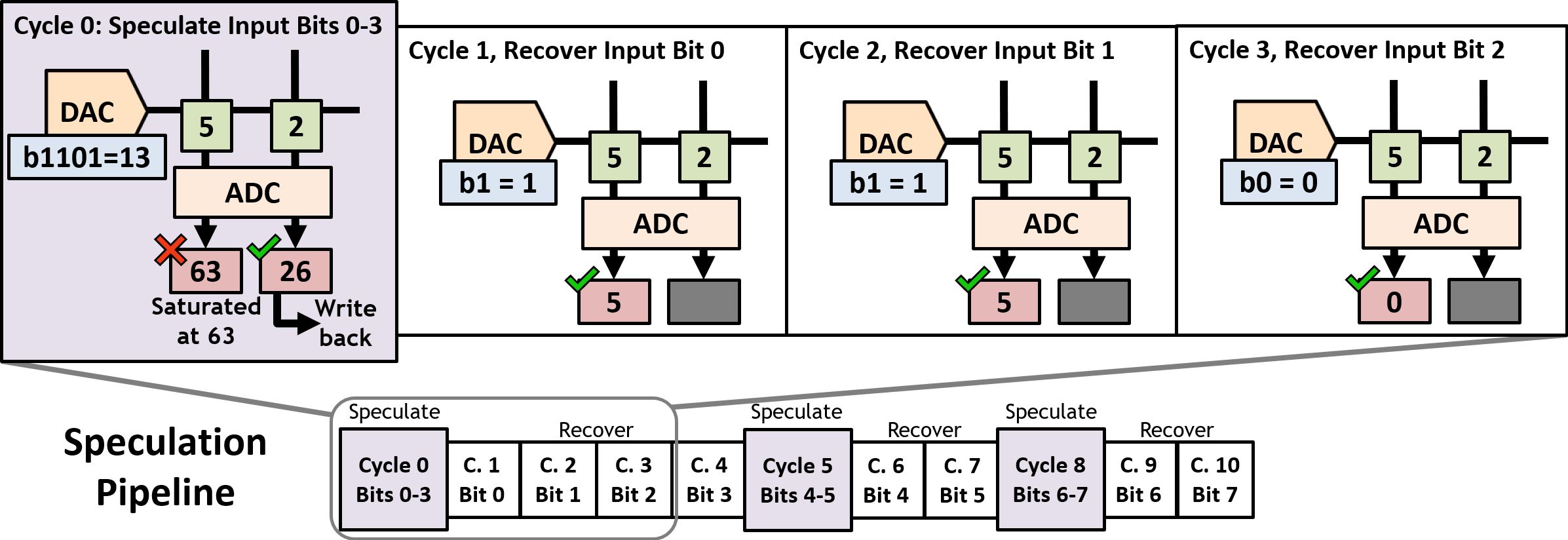}
    \centering
    \caption{Speculative Computation. Speculative cycles use 2–4 bits per input slice, with fewer ADC converts per input but a higher saturation chance. Recovery cycles use 1-bit input slices and ADCs only process columns that failed speculation.}
    \label{fig:speculation}
    \end{figure}

    The procedure for speculation and recovery is shown in Fig.~\ref{fig:speculation}. First, a 4b high-order slice is speculatively fed to the crossbar, and column sums are converted by ADCs. If a column sum is too large, it will saturate at the ADC bounds of $[-64,64)$. If an ADC output equals either of these bounds, an error is detected and marked as a speculation failure. Next, after all columns are processed, the 4b input slice is resliced into 1b slices and processed again over four recovery cycles. To save energy in recovery, ADCs are power-gated for columns that speculated successfully. In the rare event that an ADC saturates in recovery, we accept fidelity loss and propagate the saturated value. After the four recovery slices are processed, the process repeats for the following speculation and recovery cycles.

    \mysubsubsection{Dynamic Input Slicing In Hardware}
    Given a 4b DAC, analog input slices can take one of $2^{4}-1$ analog levels. For an N-bit input slice, we can use the lowest $2^{N}-1$ levels. Given a 3b \inputslice{} XXX, this corresponds to converting 0XXX. As this is only a restriction of the available range, it does not require changing the DAC hardware.
    
    To track successful/failed speculations, RAELLA stores speculation success flags in a buffer for each crossbar. In recovery, ADCs only convert column sums that failed speculation.
    
    The entire ReRAM crossbar is one unit, so all columns speculate and recover together. As it is highly likely that at least one column will fail speculation, crossbars always run recovery. Therefore, RAELLA's speculation saves energy at the cost of speed (unlike the common use of speculation for speed, \eg CPU branch prediction).
    
    Speculation also increases crossbar energy, as all columns and ReRAMs run both speculation and recovery cycles. Recovery cycles consume less energy than speculation cycles, as ReRAM devices use less energy with smaller input values~\cite{NeuroSim_Validated} and ADCs only process a small fraction of columns in recovery cycles.

    \mysubsubsection{Dynamic Input Slicing System Effects}
    RAELLA can run without speculation, processing eight recovery slices alone. With this approach, each column would require eight ADC converts for all eight input slices. With speculation, three ADC converts are needed instead to process three 2-4b speculative input slices. Across our baselines, speculation fails approximately $2\%$ of the time, requiring 2-4 recovery converts depending on which speculative slice failed. Overall, speculation succeeds $\sim{98\%}$ of the time and reduces ADC converts by $\sim{60\%}$ over a recovery-only approach. An average of three speculative converts + 0.3 recovery converts are required to process each column.
    
    While RAELLA saves ADC converts with speculation, it trades off throughput and crossbar energy. RAELLA's crossbars require eleven cycles to run all three speculation + eight recovery slices. Alternatively, a no-speculation approach could run only the eight recovery slices, increasing throughput but also increasing the number of ADC converts required.

    \mysection{RAELLA Architecture and Pipeline}
    \begin{figure}
        \includegraphics[width=\linewidth]{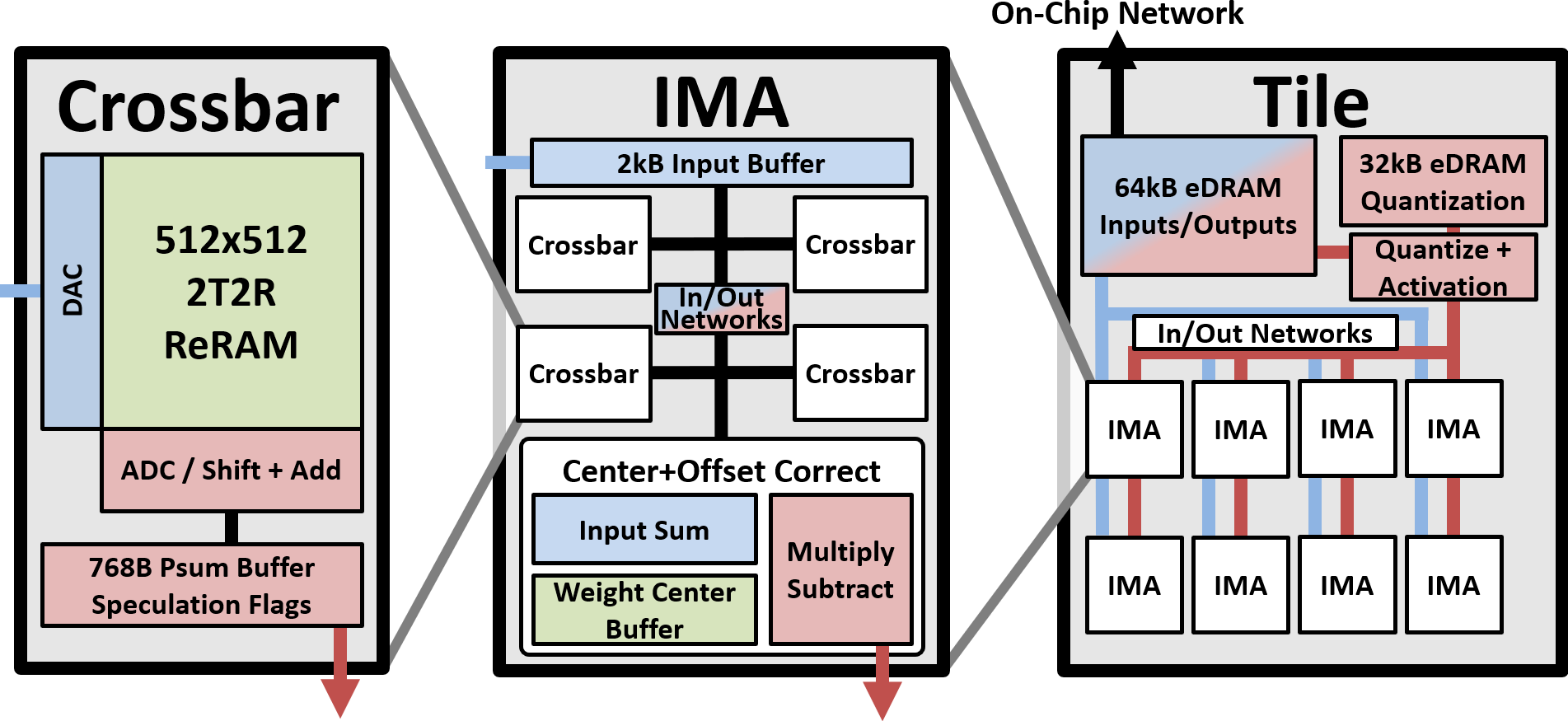}
    \centering
    \caption{The RAELLA Architecture. \numberlabel[1] The base unit is a crossbar. \numberlabel[2] Four crossbars make up an IMA. \numberlabel[3] Eight IMAs make up a tile. Components are colored blue for input storage/processing, green for weights, and red for outputs.}
    \label{fig:arch}
    \end{figure}

    The high-level RAELLA architecture is shown in Fig.~\ref{fig:arch}. RAELLA's organization mostly follows that of ISAAC~\cite{ISAAC}. We describe the RAELLA architecture from the bottom up, show RAELLA's dataflow, then describe how RAELLA reduces analog nonidealities.

    \mysubsection{Crossbar}
    Crossbars consist of $512\times512$ 2T2Rs. Each crossbar is programmed with weights from one DNN layer, and each weight filter uses 2-8 crossbar columns based on the DNN layer slicing (Section~\ref{2T2R}).
    
    To process inputs, we use 4b pulse-train DACs for their simple hardware~\cite{1024_1024_temporal_driver_pulse_train} and superior linearity~\cite{Sinangil}. Pulse-train DACs encode an N-bit \inputslice{} with a number of pulses up to $2^{N}-1$. The DAC consists of a simple row driver to apply input pulses, a 1b flip-flop to store the current input bit, and an AND gate acting as an enable signal~\cite{1024_1024_temporal_driver_pulse_train}. To output a 4b value, the most significant bit is first loaded into the flip-flop and a global clock generates eight 1ns pulses. The DAC outputs the AND'ed value of the clock and its stored value, equal to eight pulses if the bit is on and zero otherwise. Subsequent bits are loaded sequentially and run for four, two, and one pulse(s), respectively.
    
    DACs activate the 2T2R access transistors and each 2T2R device computes a \slicedproduct{} that it adds or subtracts from the \columnsum{}. \Columnsum{}s appear as a current on a column, which is buffered and scaled by a current buffer~\cite{TIMELY} before being captured as capacitor voltages and held by sample+hold circuits~\cite{sample_and_hold}.
    
    Next, four 7b ADCs~\cite{ADC} convert the 512 column sums in 100ns~\cite{ISAAC}. 7b signed ADC results are summed by shift+add circuits and accumulated in 16b psum buffers~\cite{Zhao2020LinearSQ}.

    With the most-dense slicing of two slices per weight, one crossbar may produce up to 256 psums, which are stored in a 256-entry psum buffer. Each entry stores a 16b psum + 8b success flags, for a 768B psum buffer total capacity. 
    
    In speculation/recovery cycles, inputs are streamed to crossbars once for each cycle. In speculation, ADCs process all columns. If an ADC saturates, the psum is not updated and the success flag is marked. In recovery, all success flags are checked. ADCs process and write results only for columns that failed speculation.
    
    The crossbar cycle is pipelined in two stages~\cite{ISAAC}. In the first stage, the DACs supply input pulses, the crossbar computes analog column sums, and the results are latched in sample+hold circuits. 4b pulse train DACs with 1ns/1ns on/off pulse width take 30ns to send up to 15 input pulses. Crossbars settle and produce column sums in less than a nanosecond~\cite{dot_product_engine}. In the second stage, ADCs convert sample+hold results in 100ns~\cite{ISAAC}. The overall crossbar cycle time is 100ns from the slower-stage ADC processing.

    RAELLA utilizes input bit sparsity to reduce column sum values and crossbar energy, benefiting from the high bit sparsity of unsigned inputs (Fig.~\ref{fig:bit_distributions}). If inputs are signed, RAELLA processes positive/negative inputs in two separate cycles to generate sparsity.

    \mysubsection{In-Situ Multiply Accumulate}
    Four crossbars are organized into an In-Situ Multiply Accumulate (IMA) with an input buffer~\cite{ISAAC}. An input network sends input vectors to crossbars, and if all inputs are shared between two crossbars, the input vector is multicast. To exploit temporal input reuse~\cite{efficient_processing_of_dnns, AtomLayer}, the input buffer stores reused inputs between crossbar cycles. The four crossbars can process up to $4\times512=2048$ inputs across all rows, so the buffer is sized 2kB.
    
    To support Center+Offset weights, each IMA includes a weight center buffer and digital addition circuitry to calculate input sums. A running sum is kept for each crossbar. To exploit input reuse~\cite{AtomLayer}, we add inputs to the sum when they are first used in crossbar columns and subtract when they are last used. If different crossbar columns use different subsets of the inputs, RAELLA adds/subtracts inputs in a streaming fashion while processing columns.

    \mysubsection{Tile}
    Eight IMAs are organized into a tile. Each tile includes a 64kB eDRAM buffer~\cite{ISAAC} storing 8b inputs/outputs, digital maxpool units, and quantization circuits. RAELLA digitally computes 8b per-channel quantization~\cite{Zhao2020LinearSQ}, allocating 32b per output channel to store a FP16 quantization scale and bias~\cite{Zhao2020LinearSQ}, or 32kB per tile.
    
    \mysubsection{Accelerator \& Programming}
    Like ISAAC~\cite{ISAAC}, every four tiles share a router enabling on-chip communication. When a tile completes a set of outputs, it sends data to the next tile via its router. If a layer has more weights than a tile can store, its weights are split across multiple tiles.
    
    Like other PIM accelerators~\cite{ISAAC, TIMELY, FORMS, AtomLayer}, RAELLA is programmed once for many inferences to mitigate high ReRAM write energy~\cite{NVMExplorer}. When compiling a configuration for RAELLA, we use lightweight preprocessing for Center+Offset and Adaptive Weight Slicing, as discussed in Section~\ref{Choosing weight slices}. 

\begin{figure}
    \includegraphics[width=\linewidth]{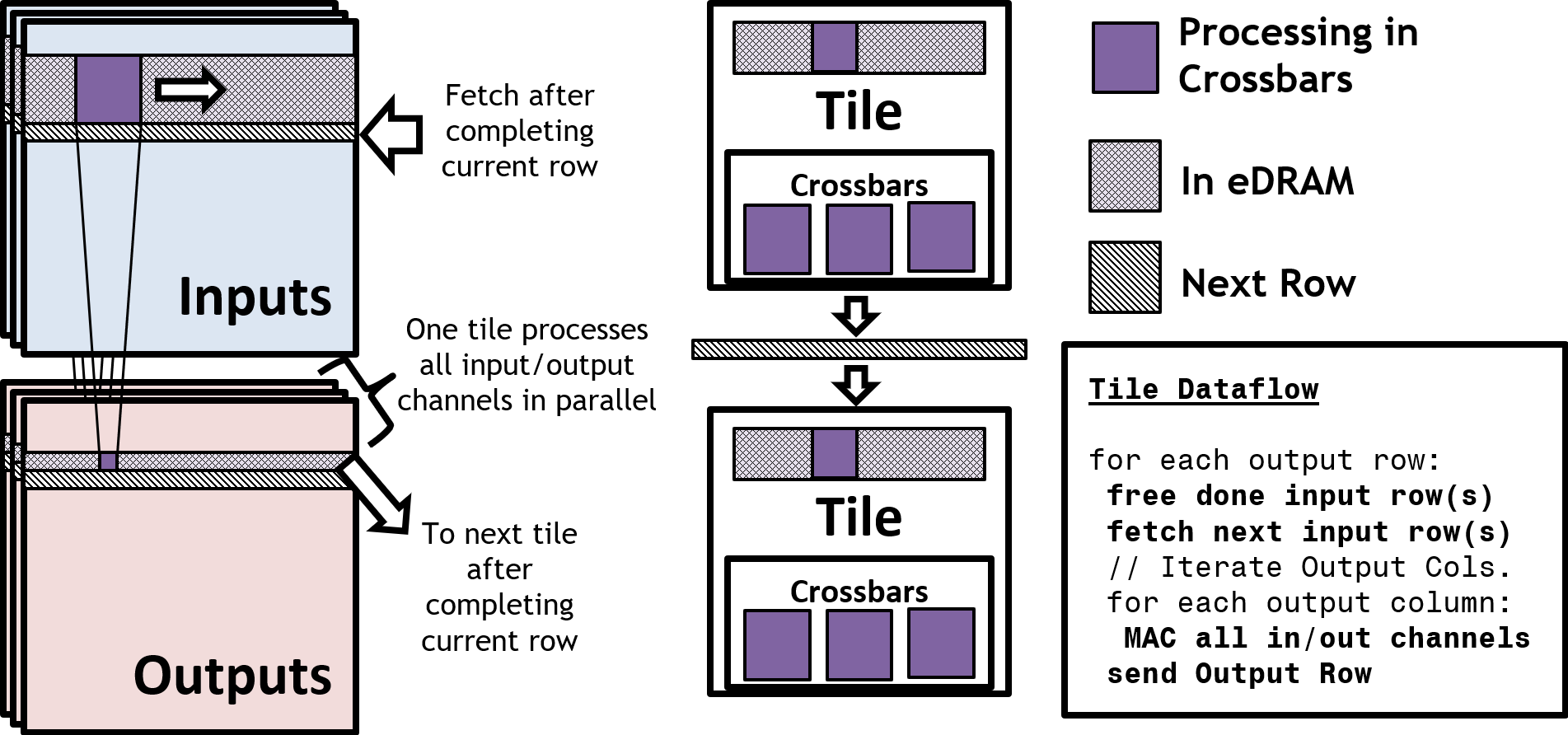}
    \centering
    \caption{Dataflow. One row of outputs for a layer are computed at a time. Tiles receive/send inputs/outputs once.}
    \label{fig:dataflow}
    \end{figure}

    \mysubsection{DNN Dataflow}
    Each DNN layer is mapped to one crossbar if it fits. Otherwise, it will spill over to more crossbars, IMAs, and tiles. RAELLA's interlayer dataflow follows ISAAC's~\cite{ISAAC} to minimize eDRAM footprint and inter-tile communication requirements. Fig.~\ref{fig:dataflow} shows RAELLA's dataflow. DNN layers are run in a pipeline across parallel tiles. Tiles generate one row of a layer's output tensor at a time, reusing previously-used input rows and fetching only new input rows. As a tile produces rows of the output tensor from top to bottom, input rows are consumed from the previous tile in the same order. Communication and data reuse patterns are coordinated by pattern generators and fixed at program time. Below the tile level, Timeloop~\cite{Timeloop} is used to find optimal data reuse patterns.
    
    RAELLA replicates weights to increase throughput following previous work~\cite{ISAAC, PipeLayer, TIMELY}. If there is space, weights are replicated in-crossbar to compute multiple convolution steps using a partial Toeplitz expansion~\cite{TIMELY, semimap}. Weights can be further replicated across crossbars, IMAs, or tiles. Replication follows a greedy scheme: while there are tiles left, the lowest-throughput layer is replicated.

    \mysubsection{RAELLA Reduces Analog Nonidealities} \label{analog_considerations}
    PIM crossbars can suffer from nonidealities such as IR drop and sneak current. RAELLA reduces these relative to ISAAC.

    High current traversing long crossbar columns causes IR drop, which can cause accuracy loss~\cite{CASCADE, SRE}. Positive/negative 2T2R devices consume current from their neighbors, reducing IR drop~\cite{2T2R_3, 1T2R_Aeris}. Furthermore, RAELLA's ADC saturates at 64, or fewer than five ReRAMs in the highest-conductance state. Therefore, RAELLA's columns must only tolerate current from five ReRAMs, compared to an ISAAC-like design that sums current for 128 ReRAMs.
    
    Sneak current, or leakage through off ReRAMs, can cause accuracy loss~\cite{CASCADE}. Sneak current is zero in 2T2R crossbars as the leakages from positive and negative ReRAMs negate~\cite{1T2R_Aeris}.
    

    \mysection{Evaluation}
    RAELLA is compared to accelerators ISAAC~\cite{ISAAC}, FORMS-8~\cite{FORMS}, and TIMELY~\cite{TIMELY}. ISAAC does not require DNN retraining. FORMS is \WeightCountLimited{} and TIMELY is \SumFidelityLimited{}, so both retrain to recover DNN accuracy.

    First, we show the efficiency and throughput gain of RAELLA in a non-retraining setting by comparing RAELLA's energy and throughput with those of ISAAC. We show that RAELLA achieves high throughput and efficiency without changing the DNN models.

    Next, we show competitiveness with DNN-retraining architectures by comparing RAELLA to FORMS and TIMELY. We show that RAELLA matches the efficiency/performance of these architectures without needing to retrain.

    Then, we show RAELLA's low accuracy loss and compare to FORMS and TIMELY. We also show the accuracy benefits of RAELLA's Center+Offset encoding.

    \mysubsection{Methodology}
    Models of RAELLA, ISAAC, and FORMS are created using Accelergy/Timeloop~\cite{accelergy,accelergy_pim,Timeloop} in the 32nm technology node. The architectures are modified to support 8b DNNs as described in Section~\ref{8b models}. Under a $600mm^{2}$ area budget, RAELLA fits 743 tiles while ISAAC and FORMS fit 1024 tiles each. Results for TIMELY are from the original paper~\cite{TIMELY}. To compare to TIMELY, we scale RAELLA to TIMELY's 65nm tech node and use TIMELY's analog components (TDC, IAdder, Charging+Comparator) and ReRAM devices~\cite{reram_device_you_use_tiox} in RAELLA. RAELLA's error budget is set to 0.09 in all tests.

    \mysubsubsection{Component Models} SRAMs are modeled in CACTI~\cite{CACTI}. Models of networks, routers, and eDRAM buffers are from ISAAC~\cite{ISAAC}. eDRAM refresh is not an issue as tiles consume data faster than a refresh period~\cite{eDRAM_refresh}. RAELLA uses the ADC~\cite{ADC} from ISAAC scaled to 7b following~\cite{ADC_scaling}. DAC, input driver, and crossbar area/energy are generated using a modified NeuroSim~\cite{NeuroSim, DNN+NeuroSim}. 2T2R area is pessimistically estimated as the sum area of two ReRAMs and two min-sized transistors, ignoring potential stacking between chip layers~\cite{4F_1T1R, stacked_MP}. DACs use a flip-flop and an AND gate for each row to generate pulse trains~\cite{1024_1024_temporal_driver_pulse_train}, where each pulse is 1ns and each 4b \inputslice{} can comprise up to 15 pulses. ReRAM parameters are taken from TIMELY~\cite{TIMELY, PRIME}, using a 0.2V read voltage and 1k$\Omega$/20k$\Omega$ on/off resistance~\cite{reram_device_you_use_tiox, dot_product_engine}. Current buffers that capture analog column sums are taken from TIMELY~\cite{TIMELY}. Outputs are quantized with a multiply/truncate and activation functions are fused into quantization~\cite{Zhao2020LinearSQ}. Maxpool units and sampling capacitors consume negligibly little energy/area~\cite{ISAAC, TIMELY}. One crossbar cycle is 100ns, and crossbars produce a set of psums every 11 cycles (three speculation \inputslice{}s + eight recovery \inputslice{}s) unless bottlenecked by the interlayer dataflow. Latency is doubled for signed inputs as positive/negative inputs are processed in separate cycles. With speculation disabled, crossbars require eight cycles and 800ns to produce a set of psums.
    
    \mysubsubsection{Models of ISAAC and FORMS} \label{8b models}
    ISAAC~\cite{ISAAC} and FORMS~\cite{FORMS} models are validated against the results presented in their papers with $<10\%$ energy and throughput error. After validating, we model ISAAC and FORMS using the same components used in RAELLA for a fair apples-to-apples comparison. In particular, the DAC/crossbars are modeled using a modified NeuroSim~\cite{NeuroSim, DNN+NeuroSim} which captures the data-dependent energy consumption of analog components. We modify both architectures and add quantization hardware to run 8b DNNs. After scaling to 8b, our ISAAC baseline has $\sim4\times$ higher efficiency and throughput than the original ISAAC while our FORMS baseline has $\sim2\times$ higher efficiency and throughput than the original FORMS. For FORMS, we use the highest reported pruning ratio. For a fair comparison, we modify ISAAC to support the partial-Toeplitz mappings~\cite{semimap, TIMELY} that RAELLA supports, which increased the throughput of ISAAC by an additional $1-1.9\times$. These mappings were not beneficial to FORMS.

    \mysubsection{DNN Models and Test Sets}
    We test on seven representative DNNs. Six are CNNs from the PyTorch~\cite{pytorch} Torchvision~\cite{torchvision} quantized library: GoogLeNet~\cite{GoogLeNet}, InceptionV3~\cite{InceptionV3}, Resnet18~\cite{ResNet}, ResNet50~\cite{ResNet}, ShuffleNetV2~\cite{shufflenet}, and MobileNetV2~\cite{MobileNetV2}. ShuffleNetV2 and MobileNetV2 are compact with small filters, while the others are large models. We report accuracy for the ImageNet~\cite{imagenet} validation set.
    
    Additionally, we test a Transformer~\cite{transformer} BERT-Large~\cite{qdqbert} on the Stanford Question Answering Dataset~\cite{SQuAD} to show RAELLA's effectiveness on cutting-edge Transformers. For BERT-Large, we accelerate the feedforward layers. Other works explore accelerating Transformer attention~\cite{analog_transformer_1, analog_transformer_2, analog_transformer_3}. BERT-Large shows RAELLA's performance with a non-ReLU activation and signed inputs.

\begin{figure*}
    \includegraphics[width=\linewidth]{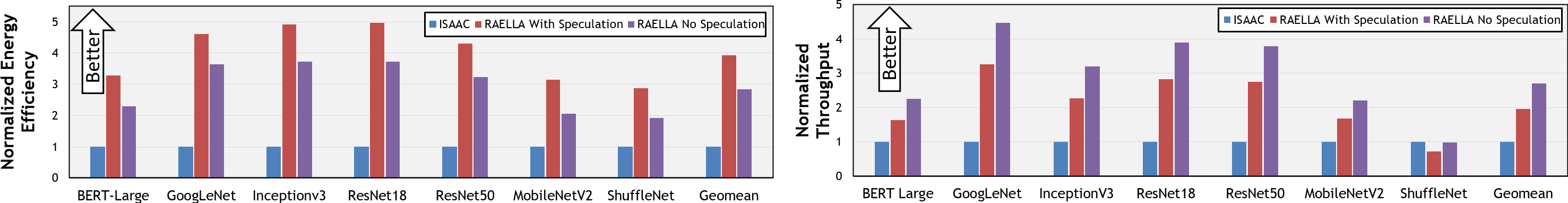}
    \centering
    \caption{Efficiency and throughput normalized to the ISAAC architecture. Both architectures run DNNs without retraining. RAELLA with/without speculation increases efficiency $3.9\times/2.8\times$ and throughput by $2.0\times/2.7\times$ geomean.}
    \label{fig:efficiency_throughput}
    \end{figure*}

    \mysubsection{Efficiency And Throughput: No Retraining}
    RAELLA is evaluated and compared to ISAAC running off-the-shelf models of all DNNs. Fig.~\ref{fig:efficiency_throughput} shows efficiency and throughput results. RAELLA improves energy efficiency $2.9$ to $4.9\times$ (geomean $3.9\times$). Efficiency gains come mainly from ADC energy reduction. RAELLA uses a 7b ADC, while ISAAC uses an 8b ADC. Furthermore, RAELLA uses larger crossbars, more bits per \inputslice{}/\weightslice{}, and speculation to reduce ADC converts by $5$ to $15\times$.

    RAELLA's throughput benefits come from large $512\times512$ (versus ISAAC's $128\times128$) and denser 2-4 bits per \weightslice{} (versus ISAAC's 2b per weight slice). Larger and denser weight storage and computation give RAELLA a throughput benefit of $0.7$ to $3.3\times$ (geomean $2.0\times$).

    Without speculation, RAELLA runs recovery slices only, reducing relative efficiency benefits to $2.8\times$ geomean due to higher ADC energy. Relative throughput increases to $2.7\times$ geomean as crossbars do not run the three speculation slices, and psums are computed in eight crossbar cycles instead of eleven.

    RAELLA is more effective with unsigned inputs and larger DNNs. Positive/negative inputs (\eg those in BERT) are processed in separate cycles, reducing throughput gains, and small filters in ShuffleNet and MobileNet poorly utilize the large crossbars of RAELLA. 

    \mysubsection{Comparison with Retraining Architectures}

    RAELLA is compared to TIMELY and FORMS-8. We show geomean ResNet18/ResNet50 results since we have data for these DNNs on all baselines. RAELLA runs off-the-shelf models, while FORMS~\cite{FORMS} runs pruned-retrained versions and TIMELY~\cite{TIMELY} runs requantized-retrained versions.

\begin{figure}
    \vspace{-9mm}
    \includegraphics[width=\linewidth]{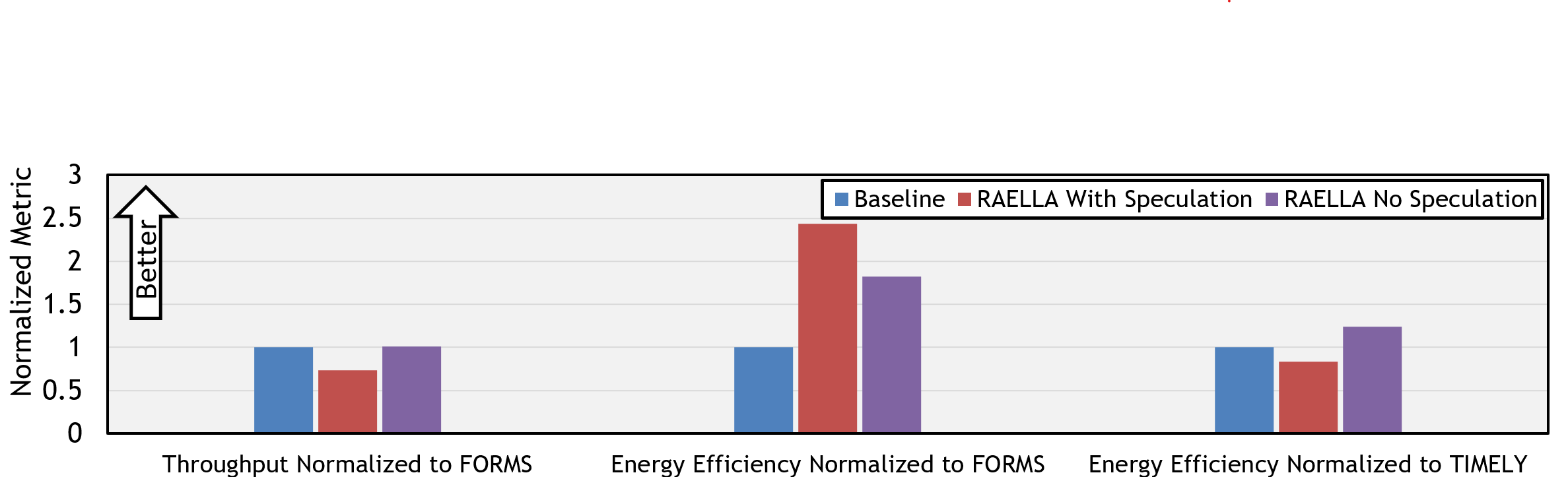}
    \centering
    \caption{Comparison with FORMS and TIMELY. FORMS and TIMELY run retrained DNNs. RAELLA offers competitive/superior throughput/efficiency without retraining.}
    \label{fig:efficiency_retraining}
    \end{figure}

    Fig.~\ref{fig:efficiency_retraining} compares RAELLA's efficiency/throughput to FORMS and TIMELY. RAELLA is able to match the throughput of FORMS and exceed the efficiency of both FORMS and TIMELY. In the TIMELY comparison, we find that 65nm RAELLA is more efficient without speculation. This is because 65nm-RAELLA uses TIMELY's analog components, including TIMELY's highly efficient ADC. Speculation is useful when ADC costs dominate, but the tradeoffs may not be worthwhile if the ADC is not a major contributor to overall energy.
    
    \mysubsection{Accuracy Comparison}
    We compare RAELLA with three baselines. RAELLA Center+Offset is the standard RAELLA, configured with a 0.09 error budget. To showcase the benefits of Center+Offset, we compare it with RAELLA Zero+Offset, which implements a common-practice differential encoding (described in Section~\ref{center_offset}) by setting centers to zero. We use the same slicings for RAELLA Center+Offset and RAELLA Zero+Offset to match efficiency/throughput. Additionally, we show the reported accuracy of FORMS and TIMELY after retraining.

    Table~\ref{tab:Accuracy Comparison} shows the accuracy results. RAELLA with Center+Offest encoding causes little to no accuracy loss. Zero+Offset (differential encoding) causes substantial accuracy degradation due to high ADC saturation rates, as described in Section~\ref{center_offset}. Zero+Offset accuracy drop varies greatly across DNNs due to varying filter weight distributions. TIMELY and FORMS recover from accuracy loss by retraining DNNs.

    \renewcommand{\arraystretch}{0}
    \begin{table}
        \centering
        \begin{tabular}{@{}lcccc@{}}
            &\shortstack{RAELLA\\Center+Offset} & \shortstack{RAELLA\\Zero+Offset} & \shortstack{FORMS\\\cite{FORMS}} & \shortstack{TIMELY\\\cite{TIMELY}} \\
            \toprule[1.5pt]
            \textbf{Retrained} & \textcolor{teal}{\textbf{No}} & \textcolor{teal}{\textbf{No}} & \textcolor{purple}{\textbf{Yes}} & \textcolor{purple}{\textbf{Yes}} \\
            \midrule
            \multicolumn{4}{@{\extracolsep{\fill}}l}{\noindent \textbf{Accuracy Drop \%. Negative is accuracy gain.}} \\
            \midrule
            ResNet18 & 0.06 & 0.16 & 0.62 & $\le0.1$ \\
            \midrule
            ResNet50 & -0.08 & 0.30 & 0.70 & $\le0.1$ \\
            \midrule
            MobileNetV2 & 0.03 & 10.17 & - & -\\
            \midrule
            ShuffleNetV2 & 0.14 & 16.36 & - & -\\
            \midrule
            GoogLeNet & -0.02 & 1.53 & - & - \\
            \midrule
            InceptionV3 & -0.03 & 3.72 & - & - \\
            \midrule
            BERT-Large & 0.12 & 0.46 & - & -\\
            \midrule
        \end{tabular}
        \caption{Accuracy Comparison. BERT-Large compares F1 loss, while others compare Imagenet Top-5 loss. Zero+Offset causes high accuracy loss; Center+Offset is essential to preserve accuracy. FORMS and TIMELY retrain, while RAELLA maintains low accuracy loss without retraining.}
        \label{tab:Accuracy Comparison}
    \end{table}

    \mysection{Ablation Studies}
    To isolate the effects of each of RAELLA's strategies, we begin with an ISAAC architecture and apply strategies sequentially. In the energy ablation, we test the efficiency benefits of each of RAELLA's strategies. In the accuracy ablation, we test RAELLA's strategies against increasing analog noise. All test setups maintain high fidelity. The four test setups are the following:
    \begin{itemize}
        \item ISAAC: an 8b ISAAC. $128\times128$ crossbars, unsigned arithmetic. Four 2b \weightslice{}s, eight 1b \inputslice{}s. 8b ADC.
        \item Center+Offset: previous setup, plus crossbar size increased to $512\times512$ 2T2R with Center+Offset arithmetic. ADC resolution is reduced to 7b.
        \item Center+Offset, Adaptive Weight Slicing: previous setup, plus weight slicings are chosen per-layer following Section~\ref{Choosing weight slices}. Most layers use three \weightslice{}s in a 4b-2b-2b pattern.
        \item RAELLA: previous setup, plus Dynamic Input Slicing and speculation enabled. RAELLA's registers/networks are added. RAELLA runs a 2-4 bit speculation \inputslice{} followed by 2-4 one-bit recovery \inputslice{}s. In recovery cycles, ADCs do not convert columns where speculation succeeded. 
    \end{itemize}

    \begin{figure}
    \includegraphics[width=\linewidth]{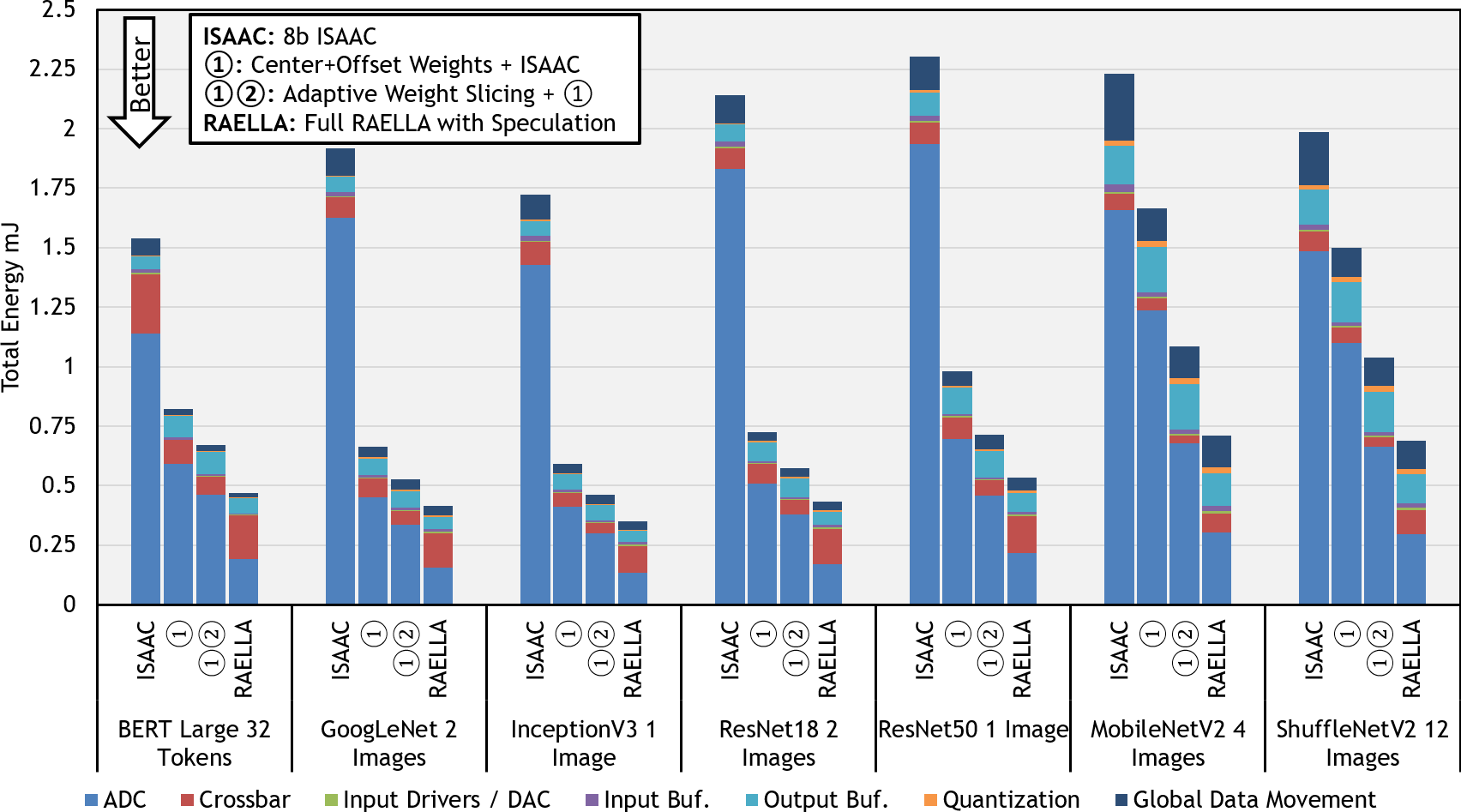}
    \centering
    \caption{Energy Ablation. Each of RAELLA's strategies increases PIM architecture efficiency. Batch size is varied across DNNs to keep overall energy in the same range.}
    \label{fig:ablation}
    \end{figure}
    \begin{figure*}
    \includegraphics[width=\linewidth]{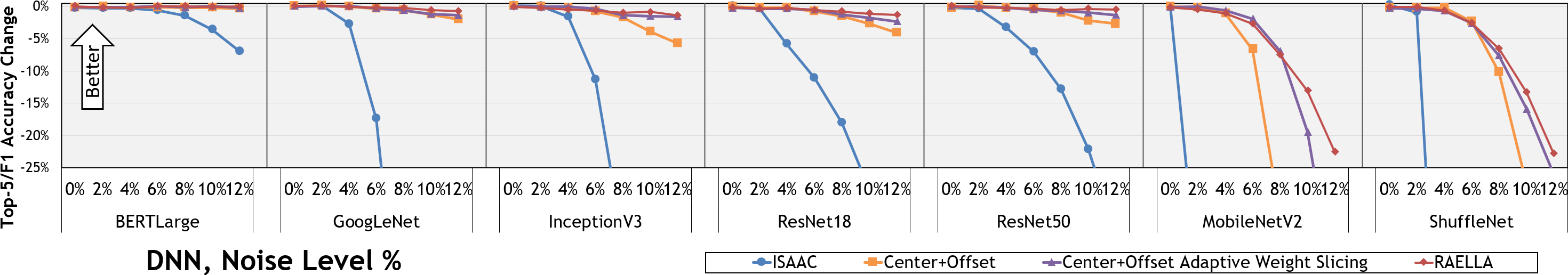}
    \centering
    \caption{Accuracy drop at increasing analog noise. Center+Offset and Adaptive Weight Slicing increase noise tolerance. Dynamic Input Slicing maintains accuracy despite speculation failures; recovery prevents accuracy loss.}
    \label{fig:accuracy}
    \vspace{-5mm}
    \end{figure*}

    \mysubsection{Energy Ablation}~\label{Energy Ablation}
    Fig.~\ref{fig:ablation} shows the following results:
    \begin{itemize}
        \item ISAAC: ADCs dominate overall energy. \emph{Converts/MAC} is $0.25$. Per-component energy breakdown varies depending on DNN input/weight values, crossbar utilization, and digital data movement requirements.
        \item Center+Offset: enables a $4\times$ scale-up in crossbar rows/columns and reduces ADC resolution. Center+Offset bit sparsity lowers crossbar energy. Large crossbars decrease data movement energy and reduce \emph{Converts/MAC} from $0.25$ to $0.063$. Digital center processing, which requires one input addition and one multiply/subtract per several hundred MACs, contributes negligible energy.
        \item Adaptive Weight Slicing: reduces ADC energy $\sim25\%$ as most layers use three weight slices instead of four. \emph{Converts/MAC} is reduced to $0.047$.
        \item Speculation: reduces ADC energy by 60\%. Increases crossbar/DAC energy due to speculation cycles. Increases the input buffer energy due to $2\times$ fetches. Usually decreases output buffer energy due to fewer psum writebacks. \emph{Converts/MAC} is $0.018$.
    \end{itemize}
    
    \mysubsection{Accuracy and In-Crossbar Noise Ablation}

    Using the same four ablation setups, we evaluate DNN accuracy running on RAELLA with varying levels of noise. All PIM architectures suffer from analog variation and noise, but RAELLA tolerates noise with lower accuracy loss.

    We model variation and noise as a Gaussian distribution that we add to column sums~\cite{dot_product_engine}. Given positive/negative sliced product sums $N_+$ and $N_-$, we model the column sum as $\mathcal{N}(\mu,\,\sigma^{2})$. For noise level $E$, we set the mean $\mu$ to the ideal column sum $(N_+-N_-)$ and the standard deviation $\sigma=E\sqrt{N_++N_-}$. After calculating a column sum, it is sent to an ADC and will be saturated at $[-64,64)$ if out of range. Noise is additive across positive/negative sliced products. We test with up to \(12\%\) error, or $\sigma\approx4$ for $512$ $2b\times2b$ MACs.

    We make two changes to ISAAC to improve noise tolerance for a fair comparison. ISAAC's encoding strategy relies on an analog circuit that sums crossbar inputs~\cite{ISAAC}. This component has been shown to degrade accuracy under noise~\cite{fidelity_encoding_exploration}, so we replace it with a digital equivalent. For BERT accuracy, we give ISAAC two cycles to process positive/negative inputs, matching RAELLA. This provides additional noise resistance. Fig.~\ref{fig:accuracy} shows the following:
    \begin{itemize}
        \item ISAAC: all DNNs suffer high accuracy loss for noise \(>4\%\). ISAAC uses unsigned weights, which have dense high-order bits (Fig.~\ref{fig:bit_distributions}). Dense bits generate larger, higher-noise values, and noise in high-order slices creates large errors in results.
        \item Center+Offset: this is critical. Offset encoding provides noise resistance~\cite{fidelity_encoding_exploration}, and Center+Offset increases bit sparsity and decreases noise. Intuitively, digital center processing moves much of the computation out of the noisy analog domain.
        \item Adaptive Weight Slicing: accuracy is further improved. RAELLA's empirical slicing strategy is noise-aware, allowing RAELLA to adapt slicing to varying levels of noise. As noise increases, Adaptive Weight Slicing uses fewer bits per weight slice to reduce error, with five weight slices for most layers at the highest tested noise.
        \item RAELLA: with speculation, RAELLA maintains accuracy similar to that of a no-speculation approach. The recovery step prevents accuracy drop due to failed speculations.
    \end{itemize}
    
    We find that RAELLA can maintain DNN accuracy at higher noise levels, while on ISAAC, all DNNs suffer sharp accuracy loss at lower noise levels. Compact DNNs suffer higher accuracy degradation from errors compared to larger DNNs~\cite{noisy_small_dnns}. BERT uniquely benefits from the sparsity generated by two-cycle positive/negative inputs. This, along with BERT's large size, allows BERT to maintain better accuracy at high noise levels.

    RAELLA can adapt to varying noise; adaptive weight slicing automatically trades off storage density and efficiency for correctness by using fewer bits per slice in higher-noise scenarios. This lets RAELLA maintain accuracy without retraining while extracting as much efficiency as possible under noise constraints.
    
    \mysection{Related Work}
    Xiao et al.~\cite{fidelity_encoding_exploration} provide an in-depth and insightful exploration of DNN accuracy versus fidelity, differential encoding, and PIM design space decisions. We urge the reader to read this work for a deeper understanding of the tradeoffs explored in RAELLA.

    Multiple works push the bounds of low ADC resolution. TinyADC~\cite{tinyadc} retrains while pruning DNN weight bits, achieving impressive reductions in \columnsum{} resolution. BRAHMS~\cite{BRAHMS} tailors ADC quantization steps for each layer to maximize DNN accuracy under fidelity loss. Guo et al.~\cite{row_oversubscription} exploit naturally-low \columnsum{}s to reduce ADC resolution and scale the number of crossbar rows used based on a column sum prediction. McDanel et al.~\cite{saturation_rram} explore low-resolution ADC quantization and DNN error tolerance. RAELLA achieves much greater ADC resolution reductions than these works (2b-3b vs. 10b).


    Newton~\cite{Newton} improves ISAAC by varying ADC resolution, using heterogeneous tiles, and using transformations that reduce computation. These are orthogonal to RAELLA; it would be interesting to see how an accelerator may combine both.

    \mysection{Conclusion}
    RAELLA shows that PIM accelerators can reduce high ADC costs without retraining or modifying DNNs. By encoding for low-resolution analog outputs and changing slicing patterns, RAELLA can reshape the distributions of computed analog values. RAELLA uses this ability to keep computed analog values low-resolution and high-fidelity while extracting as much efficiency and throughput as possible from each DNN layer. We hope that, by expanding the set of retraining-free strategies available to PIM designers, RAELLA will inspire future hardware strategies, permit novel co-design opportunities, and broaden the scope in which PIM can be used.

    \mysection{Acknowledgements}
    This work was funded in part by Ericsson, the MIT AI Hardware Program, and MIT Quest.

\bibliographystyle{ACM-Reference-Format}
\bibliography{refs.bib}


\begin{thebibliography}{82}


\ifx \showCODEN    \undefined \def \showCODEN     #1{\unskip}     \fi
\ifx \showDOI      \undefined \def \showDOI       #1{#1}\fi
\ifx \showISBNx    \undefined \def \showISBNx     #1{\unskip}     \fi
\ifx \showISBNxiii \undefined \def \showISBNxiii  #1{\unskip}     \fi
\ifx \showISSN     \undefined \def \showISSN      #1{\unskip}     \fi
\ifx \showLCCN     \undefined \def \showLCCN      #1{\unskip}     \fi
\ifx \shownote     \undefined \def \shownote      #1{#1}          \fi
\ifx \showarticletitle \undefined \def \showarticletitle #1{#1}   \fi
\ifx \showURL      \undefined \def \showURL       {\relax}        \fi
\providecommand\bibfield[2]{#2}
\providecommand\bibinfo[2]{#2}
\providecommand\natexlab[1]{#1}
\providecommand\showeprint[2][]{arXiv:#2}

\bibitem[Alibart et~al\mbox{.}(2012)]%
        {High_Precision_Programming_Variation_Intolerant}
\bibfield{author}{\bibinfo{person}{Fabien Alibart}, \bibinfo{person}{Ligang
  Gao}, \bibinfo{person}{Brian~D Hoskins}, {and} \bibinfo{person}{Dmitri~B
  Strukov}.} \bibinfo{year}{2012}\natexlab{}.
\newblock \showarticletitle{High precision tuning of state for memristive
  devices by adaptable variation-tolerant algorithm}.
\newblock \bibinfo{journal}{\emph{Nanotechnology}} \bibinfo{volume}{23},
  \bibinfo{number}{7} (\bibinfo{date}{jan} \bibinfo{year}{2012}),
  \bibinfo{pages}{075201}.
\newblock
\urldef\tempurl%
\url{https://doi.org/10.1088/0957-4484/23/7/075201}
\showDOI{\tempurl}


\bibitem[Chen et~al\mbox{.}(2017)]%
        {NeuroSim}
\bibfield{author}{\bibinfo{person}{Pai-Yu Chen}, \bibinfo{person}{Xiaochen
  Peng}, {and} \bibinfo{person}{Shimeng Yu}.} \bibinfo{year}{2017}\natexlab{}.
\newblock \showarticletitle{NeuroSim+: An integrated device-to-algorithm
  framework for benchmarking synaptic devices and array architectures}. In
  \bibinfo{booktitle}{\emph{2017 IEEE International Electron Devices Meeting
  (IEDM)}}. \bibinfo{pages}{6.1.1--6.1.4}.
\newblock
\urldef\tempurl%
\url{https://doi.org/10.1109/IEDM.2017.8268337}
\showDOI{\tempurl}


\bibitem[Chen et~al\mbox{.}(2020)]%
        {2T2R_2}
\bibfield{author}{\bibinfo{person}{Yuzong Chen}, \bibinfo{person}{Lu Lu},
  \bibinfo{person}{Bongjin Kim}, {and} \bibinfo{person}{Tony Tae-Hyoung Kim}.}
  \bibinfo{year}{2020}\natexlab{}.
\newblock \showarticletitle{Reconfigurable 2T2R ReRAM Architecture for
  Versatile Data Storage and Computing In-Memory}.
\newblock \bibinfo{journal}{\emph{IEEE Transactions on Very Large Scale
  Integration (VLSI) Systems}} \bibinfo{volume}{28}, \bibinfo{number}{12}
  (\bibinfo{year}{2020}), \bibinfo{pages}{2636--2649}.
\newblock
\urldef\tempurl%
\url{https://doi.org/10.1109/TVLSI.2020.3028848}
\showDOI{\tempurl}


\bibitem[Chen et~al\mbox{.}(2014)]%
        {DaDianNao}
\bibfield{author}{\bibinfo{person}{Yunji Chen}, \bibinfo{person}{Tao Luo},
  \bibinfo{person}{Shaoli Liu}, \bibinfo{person}{Shijin Zhang},
  \bibinfo{person}{Liqiang He}, \bibinfo{person}{Jia Wang},
  \bibinfo{person}{Ling Li}, \bibinfo{person}{Tianshi Chen},
  \bibinfo{person}{Zhiwei Xu}, \bibinfo{person}{Ninghui Sun}, {and}
  \bibinfo{person}{Olivier Temam}.} \bibinfo{year}{2014}\natexlab{}.
\newblock \showarticletitle{DaDianNao: A Machine-Learning Supercomputer}. In
  \bibinfo{booktitle}{\emph{2014 47th Annual IEEE/ACM International Symposium
  on Microarchitecture}}. \bibinfo{pages}{609--622}.
\newblock
\urldef\tempurl%
\url{https://doi.org/10.1109/MICRO.2014.58}
\showDOI{\tempurl}


\bibitem[Chi et~al\mbox{.}(2016)]%
        {PRIME}
\bibfield{author}{\bibinfo{person}{Ping Chi}, \bibinfo{person}{Shuangchen Li},
  \bibinfo{person}{Cong Xu}, \bibinfo{person}{Tao Zhang},
  \bibinfo{person}{Jishen Zhao}, \bibinfo{person}{Yongpan Liu},
  \bibinfo{person}{Yu Wang}, {and} \bibinfo{person}{Yuan Xie}.}
  \bibinfo{year}{2016}\natexlab{}.
\newblock \showarticletitle{PRIME: A Novel Processing-in-Memory Architecture
  for Neural Network Computation in ReRAM-Based Main Memory}. In
  \bibinfo{booktitle}{\emph{2016 ACM/IEEE 43rd Annual International Symposium
  on Computer Architecture (ISCA)}}. \bibinfo{pages}{27--39}.
\newblock
\urldef\tempurl%
\url{https://doi.org/10.1109/ISCA.2016.13}
\showDOI{\tempurl}


\bibitem[Choi et~al\mbox{.}(2019)]%
        {ibm_low_precision_inference}
\bibfield{author}{\bibinfo{person}{Jungwook Choi}, \bibinfo{person}{Swagath
  Venkataramani}, \bibinfo{person}{Vijayalakshmi~(Viji) Srinivasan},
  \bibinfo{person}{Kailash Gopalakrishnan}, \bibinfo{person}{Zhuo Wang}, {and}
  \bibinfo{person}{Pierce Chuang}.} \bibinfo{year}{2019}\natexlab{}.
\newblock \showarticletitle{Accurate and Efficient 2-bit Quantized Neural
  Networks}. In \bibinfo{booktitle}{\emph{Proceedings of Machine Learning and
  Systems}}, \bibfield{editor}{\bibinfo{person}{A.~Talwalkar},
  \bibinfo{person}{V.~Smith}, {and} \bibinfo{person}{M.~Zaharia}} (Eds.),
  Vol.~\bibinfo{volume}{1}. \bibinfo{pages}{348--359}.
\newblock
\urldef\tempurl%
\url{https://proceedings.mlsys.org/paper/2019/file/006f52e9102a8d3be2fe5614f42ba989-Paper.pdf}
\showURL{%
\tempurl}


\bibitem[Chou et~al\mbox{.}(2019)]%
        {CASCADE}
\bibfield{author}{\bibinfo{person}{Teyuh Chou}, \bibinfo{person}{Wei Tang},
  \bibinfo{person}{Jacob Botimer}, {and} \bibinfo{person}{Zhengya Zhang}.}
  \bibinfo{year}{2019}\natexlab{}.
\newblock \showarticletitle{CASCADE: Connecting RRAMs to Extend Analog Dataflow
  In An End-To-End In-Memory Processing Paradigm}. In
  \bibinfo{booktitle}{\emph{Proceedings of the 52nd Annual IEEE/ACM
  International Symposium on Microarchitecture}} (Columbus, OH, USA)
  \emph{(\bibinfo{series}{MICRO '52})}. \bibinfo{publisher}{Association for
  Computing Machinery}, \bibinfo{address}{New York, NY, USA},
  \bibinfo{pages}{114–125}.
\newblock
\showISBNx{9781450369381}
\urldef\tempurl%
\url{https://doi.org/10.1145/3352460.3358328}
\showDOI{\tempurl}


\bibitem[Chu et~al\mbox{.}(2020)]%
        {PIM-Prune}
\bibfield{author}{\bibinfo{person}{Chaoqun Chu}, \bibinfo{person}{Yanzhi Wang},
  \bibinfo{person}{Yilong Zhao}, \bibinfo{person}{Xiaolong Ma},
  \bibinfo{person}{Shaokai Ye}, \bibinfo{person}{Yunyan Hong},
  \bibinfo{person}{Xiaoyao Liang}, \bibinfo{person}{Yinhe Han}, {and}
  \bibinfo{person}{Li Jiang}.} \bibinfo{year}{2020}\natexlab{}.
\newblock \showarticletitle{PIM-Prune: Fine-Grain DCNN Pruning for
  Crossbar-Based Process-In-Memory Architecture}. In
  \bibinfo{booktitle}{\emph{2020 57th ACM/IEEE Design Automation Conference
  (DAC)}}. \bibinfo{pages}{1--6}.
\newblock
\urldef\tempurl%
\url{https://doi.org/10.1109/DAC18072.2020.9218523}
\showDOI{\tempurl}


\bibitem[Colangelo et~al\mbox{.}(2018)]%
        {intel_low_precision_dnn}
\bibfield{author}{\bibinfo{person}{Philip Colangelo}, \bibinfo{person}{Nasibeh
  Nasiri}, \bibinfo{person}{Eriko Nurvitadhi}, \bibinfo{person}{Asit Mishra},
  \bibinfo{person}{Martin Margala}, {and} \bibinfo{person}{Kevin Nealis}.}
  \bibinfo{year}{2018}\natexlab{}.
\newblock \showarticletitle{Exploration of Low Numeric Precision Deep Learning
  Inference Using Intel® FPGAs}. In \bibinfo{booktitle}{\emph{2018 IEEE 26th
  Annual International Symposium on Field-Programmable Custom Computing
  Machines (FCCM)}}. \bibinfo{pages}{73--80}.
\newblock
\urldef\tempurl%
\url{https://doi.org/10.1109/FCCM.2018.00020}
\showDOI{\tempurl}


\bibitem[Deng et~al\mbox{.}(2009)]%
        {imagenet}
\bibfield{author}{\bibinfo{person}{Jia Deng}, \bibinfo{person}{Wei Dong},
  \bibinfo{person}{Richard Socher}, \bibinfo{person}{Li-Jia Li},
  \bibinfo{person}{Kai Li}, {and} \bibinfo{person}{Li Fei-Fei}.}
  \bibinfo{year}{2009}\natexlab{}.
\newblock \showarticletitle{ImageNet: A large-scale hierarchical image
  database}. In \bibinfo{booktitle}{\emph{2009 IEEE Conference on Computer
  Vision and Pattern Recognition}}. \bibinfo{pages}{248--255}.
\newblock
\urldef\tempurl%
\url{https://doi.org/10.1109/CVPR.2009.5206848}
\showDOI{\tempurl}


\bibitem[Deng et~al\mbox{.}(2020)]%
        {semimap}
\bibfield{author}{\bibinfo{person}{Lei Deng}, \bibinfo{person}{Ling Liang},
  \bibinfo{person}{Guanrui Wang}, \bibinfo{person}{Liang Chang},
  \bibinfo{person}{Xing Hu}, \bibinfo{person}{Xin Ma}, \bibinfo{person}{Liu
  Liu}, \bibinfo{person}{Jing Pei}, \bibinfo{person}{Guoqi Li}, {and}
  \bibinfo{person}{Yuan Xie}.} \bibinfo{year}{2020}\natexlab{}.
\newblock \showarticletitle{SemiMap: A Semi-Folded Convolution Mapping for
  Speed-Overhead Balance on Crossbars}.
\newblock \bibinfo{journal}{\emph{IEEE Transactions on Computer-Aided Design of
  Integrated Circuits and Systems}} \bibinfo{volume}{39}, \bibinfo{number}{1}
  (\bibinfo{year}{2020}), \bibinfo{pages}{117--130}.
\newblock
\urldef\tempurl%
\url{https://doi.org/10.1109/TCAD.2018.2883959}
\showDOI{\tempurl}


\bibitem[Fasoli et~al\mbox{.}(2021)]%
        {low_precision_rnn}
\bibfield{author}{\bibinfo{person}{Andrea Fasoli}, \bibinfo{person}{Chia-Yu
  Chen}, \bibinfo{person}{Mauricio Serrano}, \bibinfo{person}{Xiao Sun},
  \bibinfo{person}{Naigang Wang}, \bibinfo{person}{Swagath Venkataramani},
  \bibinfo{person}{George Saon}, \bibinfo{person}{Xiaodong Cui},
  \bibinfo{person}{Brian Kingsbury}, \bibinfo{person}{Wei Zhang},
  \bibinfo{person}{Zoltán Tüske}, {and} \bibinfo{person}{Kailash
  Gopalakrishnan}.} \bibinfo{year}{2021}\natexlab{}.
\newblock \showarticletitle{{4-Bit Quantization of LSTM-Based Speech
  Recognition Models}}. In \bibinfo{booktitle}{\emph{Proc. Interspeech 2021}}.
  \bibinfo{pages}{2586--2590}.
\newblock
\urldef\tempurl%
\url{https://doi.org/10.21437/Interspeech.2021-1962}
\showDOI{\tempurl}


\bibitem[Gao et~al\mbox{.}(2013)]%
        {reram_device_you_use_tiox}
\bibfield{author}{\bibinfo{person}{Ligang Gao}, \bibinfo{person}{Fabien
  Alibart}, {and} \bibinfo{person}{Dmitri~B. Strukov}.}
  \bibinfo{year}{2013}\natexlab{}.
\newblock \showarticletitle{A High Resolution Nonvolatile Analog Memory Ionic
  Devices}.
\newblock


\bibitem[Gonugondla et~al\mbox{.}(2020)]%
        {fundamental_limits_of_crossbar_precision}
\bibfield{author}{\bibinfo{person}{Sujan~K. Gonugondla},
  \bibinfo{person}{Charbel Sakr}, \bibinfo{person}{Hassan Dbouk}, {and}
  \bibinfo{person}{Naresh~R. Shanbhag}.} \bibinfo{year}{2020}\natexlab{}.
\newblock \showarticletitle{Fundamental Limits on the Precision of In-Memory
  Architectures}. In \bibinfo{booktitle}{\emph{Proceedings of the 39th
  International Conference on Computer-Aided Design}} (Virtual Event, USA)
  \emph{(\bibinfo{series}{ICCAD '20})}. \bibinfo{publisher}{Association for
  Computing Machinery}, \bibinfo{address}{New York, NY, USA}, Article
  \bibinfo{articleno}{128}, \bibinfo{numpages}{9}~pages.
\newblock
\showISBNx{9781450380263}
\urldef\tempurl%
\url{https://doi.org/10.1145/3400302.3416344}
\showDOI{\tempurl}


\bibitem[Guo et~al\mbox{.}(2022)]%
        {row_oversubscription}
\bibfield{author}{\bibinfo{person}{Mengyu Guo}, \bibinfo{person}{Zihan Zhang},
  \bibinfo{person}{Jianfei Jiang}, \bibinfo{person}{Qin Wang}, {and}
  \bibinfo{person}{Naifeng Jing}.} \bibinfo{year}{2022}\natexlab{}.
\newblock \showarticletitle{Boosting ReRAM-based DNN by Row Activation
  Oversubscription}. In \bibinfo{booktitle}{\emph{2022 27th Asia and South
  Pacific Design Automation Conference (ASP-DAC)}}. \bibinfo{pages}{604--609}.
\newblock
\urldef\tempurl%
\url{https://doi.org/10.1109/ASP-DAC52403.2022.9712520}
\showDOI{\tempurl}


\bibitem[He et~al\mbox{.}(2016)]%
        {ResNet}
\bibfield{author}{\bibinfo{person}{Kaiming He}, \bibinfo{person}{X. Zhang},
  \bibinfo{person}{Shaoqing Ren}, {and} \bibinfo{person}{Jian Sun}.}
  \bibinfo{year}{2016}\natexlab{}.
\newblock \showarticletitle{Deep Residual Learning for Image Recognition}.
\newblock \bibinfo{journal}{\emph{2016 IEEE Conference on Computer Vision and
  Pattern Recognition (CVPR)}} (\bibinfo{year}{2016}),
  \bibinfo{pages}{770--778}.
\newblock


\bibitem[Hu et~al\mbox{.}(2016)]%
        {dot_product_engine}
\bibfield{author}{\bibinfo{person}{Miao Hu}, \bibinfo{person}{John~Paul
  Strachan}, \bibinfo{person}{Zhiyong Li}, \bibinfo{person}{Emmanuelle~M.
  Grafals}, \bibinfo{person}{Noraica Davila}, \bibinfo{person}{Catherine
  Graves}, \bibinfo{person}{Sity Lam}, \bibinfo{person}{Ning Ge},
  \bibinfo{person}{Jianhua~Joshua Yang}, {and} \bibinfo{person}{R.~Stanley
  Williams}.} \bibinfo{year}{2016}\natexlab{}.
\newblock \showarticletitle{Dot-product engine for neuromorphic computing:
  Programming 1T1M crossbar to accelerate matrix-vector multiplication}. In
  \bibinfo{booktitle}{\emph{2016 53nd ACM/EDAC/IEEE Design Automation
  Conference (DAC)}}. \bibinfo{pages}{1--6}.
\newblock
\urldef\tempurl%
\url{https://doi.org/10.1145/2897937.2898010}
\showDOI{\tempurl}


\bibitem[Jouppi et~al\mbox{.}(2015)]%
        {CACTI}
\bibfield{author}{\bibinfo{person}{Norman~P. Jouppi},
  \bibinfo{person}{Andrew~B. Kahng}, \bibinfo{person}{Naveen Muralimanohar},
  {and} \bibinfo{person}{Vaishnav Srinivas}.} \bibinfo{year}{2015}\natexlab{}.
\newblock \showarticletitle{CACTI-IO: CACTI With OFF-Chip Power-Area-Timing
  Models}.
\newblock \bibinfo{journal}{\emph{IEEE Transactions on Very Large Scale
  Integration (VLSI) Systems}} \bibinfo{volume}{23}, \bibinfo{number}{7}
  (\bibinfo{year}{2015}), \bibinfo{pages}{1254--1267}.
\newblock
\urldef\tempurl%
\url{https://doi.org/10.1109/TVLSI.2014.2334635}
\showDOI{\tempurl}


\bibitem[Joy et~al\mbox{.}(2016)]%
        {hyperparameter_tuning}
\bibfield{author}{\bibinfo{person}{Tinu~Theckel Joy}, \bibinfo{person}{Santu
  Rana}, \bibinfo{person}{Sunil Gupta}, {and} \bibinfo{person}{Svetha
  Venkatesh}.} \bibinfo{year}{2016}\natexlab{}.
\newblock \showarticletitle{Hyperparameter tuning for big data using Bayesian
  optimisation}. In \bibinfo{booktitle}{\emph{2016 23rd International
  Conference on Pattern Recognition (ICPR)}}. \bibinfo{pages}{2574--2579}.
\newblock
\urldef\tempurl%
\url{https://doi.org/10.1109/ICPR.2016.7900023}
\showDOI{\tempurl}


\bibitem[Kim et~al\mbox{.}(2022)]%
        {sram_analog_sub}
\bibfield{author}{\bibinfo{person}{Sangyeob Kim}, \bibinfo{person}{Sangjin
  Kim}, \bibinfo{person}{Soyeon Um}, \bibinfo{person}{Soyeon Kim},
  \bibinfo{person}{Kwantae Kim}, {and} \bibinfo{person}{Hoi-Jun Yoo}.}
  \bibinfo{year}{2022}\natexlab{}.
\newblock \showarticletitle{Neuro-CIM: A 310.4 TOPS/W Neuromorphic
  Computing-in-Memory Processor with Low WL/BL activity and Digital-Analog
  Mixed-mode Neuron Firing}. In \bibinfo{booktitle}{\emph{2022 IEEE Symposium
  on VLSI Technology and Circuits (VLSI Technology and Circuits)}}.
  \bibinfo{pages}{38--39}.
\newblock
\urldef\tempurl%
\url{https://doi.org/10.1109/VLSITechnologyandCir46769.2022.9830276}
\showDOI{\tempurl}


\bibitem[Klachko et~al\mbox{.}(2019)]%
        {noise_tolerance}
\bibfield{author}{\bibinfo{person}{Michael Klachko},
  \bibinfo{person}{Mohammad~Reza Mahmoodi}, {and} \bibinfo{person}{Dmitri
  Strukov}.} \bibinfo{year}{2019}\natexlab{}.
\newblock \showarticletitle{Improving Noise Tolerance of Mixed-Signal Neural
  Networks}. In \bibinfo{booktitle}{\emph{2019 International Joint Conference
  on Neural Networks (IJCNN)}}. \bibinfo{pages}{1--8}.
\newblock
\urldef\tempurl%
\url{https://doi.org/10.1109/IJCNN.2019.8851966}
\showDOI{\tempurl}


\bibitem[Krishnamoorthi(2018)]%
        {quant_whitepaper}
\bibfield{author}{\bibinfo{person}{Raghuraman Krishnamoorthi}.}
  \bibinfo{year}{2018}\natexlab{}.
\newblock \showarticletitle{Quantizing deep convolutional networks for
  efficient inference: {A} whitepaper}.
\newblock \bibinfo{journal}{\emph{CoRR}}  \bibinfo{volume}{abs/1806.08342}
  (\bibinfo{year}{2018}).
\newblock
\showeprint[arXiv]{1806.08342}
\urldef\tempurl%
\url{http://arxiv.org/abs/1806.08342}
\showURL{%
\tempurl}


\bibitem[Kull et~al\mbox{.}(2013)]%
        {ADC}
\bibfield{author}{\bibinfo{person}{Lukas Kull}, \bibinfo{person}{Thomas Toifl},
  \bibinfo{person}{Martin Schmatz}, \bibinfo{person}{Pier~Andrea Francese},
  \bibinfo{person}{Christian Menolfi}, \bibinfo{person}{Matthias Braendli},
  \bibinfo{person}{Marcel Kossel}, \bibinfo{person}{Thomas Morf},
  \bibinfo{person}{Toke~Meyer Andersen}, {and} \bibinfo{person}{Yusuf
  Leblebici}.} \bibinfo{year}{2013}\natexlab{}.
\newblock \showarticletitle{A 3.1mW 8b 1.2GS/s single-channel asynchronous SAR
  ADC with alternate comparators for enhanced speed in 32nm digital SOI CMOS}.
  In \bibinfo{booktitle}{\emph{2013 IEEE International Solid-State Circuits
  Conference Digest of Technical Papers}}. \bibinfo{pages}{468--469}.
\newblock
\urldef\tempurl%
\url{https://doi.org/10.1109/ISSCC.2013.6487818}
\showDOI{\tempurl}


\bibitem[Li et~al\mbox{.}(2020)]%
        {TIMELY}
\bibfield{author}{\bibinfo{person}{Weitao Li}, \bibinfo{person}{Pengfei Xu},
  \bibinfo{person}{Yang Zhao}, \bibinfo{person}{Haitong Li},
  \bibinfo{person}{Yuan Xie}, {and} \bibinfo{person}{Yingyan Lin}.}
  \bibinfo{year}{2020}\natexlab{}.
\newblock \showarticletitle{TIMELY: Pushing Data Movements and Interfaces in
  PIM Accelerators towards Local and in Time Domain}. In
  \bibinfo{booktitle}{\emph{Proceedings of the ACM/IEEE 47th Annual
  International Symposium on Computer Architecture}} (Virtual Event)
  \emph{(\bibinfo{series}{ISCA '20})}. \bibinfo{publisher}{IEEE Press},
  \bibinfo{pages}{832–845}.
\newblock
\showISBNx{9781728146614}
\urldef\tempurl%
\url{https://doi.org/10.1109/ISCA45697.2020.00073}
\showDOI{\tempurl}


\bibitem[Liang et~al\mbox{.}(2021)]%
        {Pruning_Quant_Survey}
\bibfield{author}{\bibinfo{person}{Tailin Liang}, \bibinfo{person}{John
  Glossner}, \bibinfo{person}{Lei Wang}, \bibinfo{person}{Shaobo Shi}, {and}
  \bibinfo{person}{Xiaotong Zhang}.} \bibinfo{year}{2021}\natexlab{}.
\newblock \showarticletitle{Pruning and quantization for deep neural network
  acceleration: A survey}.
\newblock \bibinfo{journal}{\emph{Neurocomputing}}  \bibinfo{volume}{461}
  (\bibinfo{year}{2021}), \bibinfo{pages}{370--403}.
\newblock
\showISSN{0925-2312}
\urldef\tempurl%
\url{https://doi.org/10.1016/j.neucom.2021.07.045}
\showDOI{\tempurl}


\bibitem[Lin et~al\mbox{.}(2019)]%
        {learning_sparsity_for_reram}
\bibfield{author}{\bibinfo{person}{Jilan Lin}, \bibinfo{person}{Zhenhua Zhu},
  \bibinfo{person}{Yu Wang}, {and} \bibinfo{person}{Yuan Xie}.}
  \bibinfo{year}{2019}\natexlab{}.
\newblock \showarticletitle{Learning the Sparsity for ReRAM: Mapping and
  Pruning Sparse Neural Network for ReRAM Based Accelerator}. In
  \bibinfo{booktitle}{\emph{Proceedings of the 24th Asia and South Pacific
  Design Automation Conference}} (Tokyo, Japan) \emph{(\bibinfo{series}{ASPDAC
  '19})}. \bibinfo{publisher}{Association for Computing Machinery},
  \bibinfo{address}{New York, NY, USA}, \bibinfo{pages}{639–644}.
\newblock
\showISBNx{9781450360074}
\urldef\tempurl%
\url{https://doi.org/10.1145/3287624.3287715}
\showDOI{\tempurl}


\bibitem[Linn et~al\mbox{.}(2010)]%
        {stacked_MP}
\bibfield{author}{\bibinfo{person}{Eike Linn}, \bibinfo{person}{Roland
  Rosezin}, \bibinfo{person}{Carsten K{\"u}geler}, {and}
  \bibinfo{person}{Rainer Waser}.} \bibinfo{year}{2010}\natexlab{}.
\newblock \showarticletitle{Complementary resistive switches for passive
  nanocrossbar memories}.
\newblock \bibinfo{journal}{\emph{Nature Materials}} \bibinfo{volume}{9},
  \bibinfo{number}{5} (\bibinfo{date}{01 May} \bibinfo{year}{2010}),
  \bibinfo{pages}{403--406}.
\newblock
\showISSN{1476-4660}
\urldef\tempurl%
\url{https://doi.org/10.1038/nmat2748}
\showDOI{\tempurl}


\bibitem[Liu et~al\mbox{.}(2020)]%
        {2T2R_3}
\bibfield{author}{\bibinfo{person}{Qi Liu}, \bibinfo{person}{Bin Gao},
  \bibinfo{person}{Peng Yao}, \bibinfo{person}{Dong Wu},
  \bibinfo{person}{Junren Chen}, \bibinfo{person}{Yachuan Pang},
  \bibinfo{person}{Wenqiang Zhang}, \bibinfo{person}{Yan Liao},
  \bibinfo{person}{Cheng-Xin Xue}, \bibinfo{person}{Wei-Hao Chen},
  \bibinfo{person}{Jianshi Tang}, \bibinfo{person}{Yu Wang},
  \bibinfo{person}{Meng-Fan Chang}, \bibinfo{person}{He Qian}, {and}
  \bibinfo{person}{Huaqiang Wu}.} \bibinfo{year}{2020}\natexlab{}.
\newblock \showarticletitle{33.2 A Fully Integrated Analog ReRAM Based
  78.4TOPS/W Compute-In-Memory Chip with Fully Parallel MAC Computing}. In
  \bibinfo{booktitle}{\emph{2020 IEEE International Solid- State Circuits
  Conference - (ISSCC)}}. \bibinfo{pages}{500--502}.
\newblock
\urldef\tempurl%
\url{https://doi.org/10.1109/ISSCC19947.2020.9062953}
\showDOI{\tempurl}


\bibitem[Lu et~al\mbox{.}(2021)]%
        {NeuroSim_Validated}
\bibfield{author}{\bibinfo{person}{Anni Lu}, \bibinfo{person}{Xiaochen Peng},
  \bibinfo{person}{Wantong Li}, \bibinfo{person}{Hongwu Jiang}, {and}
  \bibinfo{person}{Shimeng Yu}.} \bibinfo{year}{2021}\natexlab{}.
\newblock \showarticletitle{NeuroSim Validation with 40nm RRAM
  Compute-in-Memory Macro}. In \bibinfo{booktitle}{\emph{2021 IEEE 3rd
  International Conference on Artificial Intelligence Circuits and Systems
  (AICAS)}}. \bibinfo{pages}{1--4}.
\newblock
\urldef\tempurl%
\url{https://doi.org/10.1109/AICAS51828.2021.9458501}
\showDOI{\tempurl}


\bibitem[Ma et~al\mbox{.}(2018)]%
        {shufflenet}
\bibfield{author}{\bibinfo{person}{Ningning Ma}, \bibinfo{person}{Xiangyu
  Zhang}, \bibinfo{person}{Hai-Tao Zheng}, {and} \bibinfo{person}{Jian Sun}.}
  \bibinfo{year}{2018}\natexlab{}.
\newblock \showarticletitle{ShuffleNet V2: Practical Guidelines for Efficient
  CNN Architecture Design}. In \bibinfo{booktitle}{\emph{Proceedings of the
  European Conference on Computer Vision (ECCV)}}.
\newblock


\bibitem[Marcel and Rodriguez(2010)]%
        {torchvision}
\bibfield{author}{\bibinfo{person}{S\'{e}bastien Marcel} {and}
  \bibinfo{person}{Yann Rodriguez}.} \bibinfo{year}{2010}\natexlab{}.
\newblock \showarticletitle{Torchvision the Machine-Vision Package of Torch}.
  In \bibinfo{booktitle}{\emph{Proceedings of the 18th ACM International
  Conference on Multimedia}} (Firenze, Italy) \emph{(\bibinfo{series}{MM
  '10})}. \bibinfo{publisher}{Association for Computing Machinery},
  \bibinfo{address}{New York, NY, USA}, \bibinfo{pages}{1485–1488}.
\newblock
\showISBNx{9781605589336}
\urldef\tempurl%
\url{https://doi.org/10.1145/1873951.1874254}
\showDOI{\tempurl}


\bibitem[Marinella et~al\mbox{.}(2018)]%
        {1024_1024_temporal_driver_pulse_train}
\bibfield{author}{\bibinfo{person}{Matthew~J. Marinella},
  \bibinfo{person}{Sapan Agarwal}, \bibinfo{person}{Alexander Hsia},
  \bibinfo{person}{Isaac Richter}, \bibinfo{person}{Robin Jacobs-Gedrim},
  \bibinfo{person}{John Niroula}, \bibinfo{person}{Steven~J. Plimpton},
  \bibinfo{person}{Engin Ipek}, {and} \bibinfo{person}{Conrad~D. James}.}
  \bibinfo{year}{2018}\natexlab{}.
\newblock \showarticletitle{Multiscale Co-Design Analysis of Energy, Latency,
  Area, and Accuracy of a ReRAM Analog Neural Training Accelerator}.
\newblock \bibinfo{journal}{\emph{IEEE Journal on Emerging and Selected Topics
  in Circuits and Systems}} \bibinfo{volume}{8}, \bibinfo{number}{1}
  (\bibinfo{year}{2018}), \bibinfo{pages}{86--101}.
\newblock
\urldef\tempurl%
\url{https://doi.org/10.1109/JETCAS.2018.2796379}
\showDOI{\tempurl}


\bibitem[McDanel et~al\mbox{.}(2021)]%
        {saturation_rram}
\bibfield{author}{\bibinfo{person}{Bradley McDanel}, \bibinfo{person}{Sai~Qian
  Zhang}, {and} \bibinfo{person}{H.~T. Kung}.} \bibinfo{year}{2021}\natexlab{}.
\newblock \showarticletitle{Saturation RRAM Leveraging Bit-Level Sparsity
  Resulting from Term Quantization}. In \bibinfo{booktitle}{\emph{2021 IEEE
  International Symposium on Circuits and Systems (ISCAS)}}.
  \bibinfo{pages}{1--5}.
\newblock
\urldef\tempurl%
\url{https://doi.org/10.1109/ISCAS51556.2021.9401293}
\showDOI{\tempurl}


\bibitem[Mittal(2019)]%
        {reram_survey}
\bibfield{author}{\bibinfo{person}{Sparsh Mittal}.}
  \bibinfo{year}{2019}\natexlab{}.
\newblock \showarticletitle{A Survey of ReRAM-Based Architectures for
  Processing-In-Memory and Neural Networks}.
\newblock \bibinfo{journal}{\emph{Machine Learning and Knowledge Extraction}}
  \bibinfo{volume}{1}, \bibinfo{number}{1} (\bibinfo{year}{2019}),
  \bibinfo{pages}{75--114}.
\newblock
\showISSN{2504-4990}
\urldef\tempurl%
\url{https://doi.org/10.3390/make1010005}
\showDOI{\tempurl}


\bibitem[Murmann(2013)]%
        {ADC_efficiency_max}
\bibfield{author}{\bibinfo{person}{Boris Murmann}.}
  \bibinfo{year}{2013}\natexlab{}.
\newblock \showarticletitle{Energy limits in A/D converters}. In
  \bibinfo{booktitle}{\emph{2013 IEEE Faible Tension Faible Consommation}}.
  \bibinfo{pages}{1--4}.
\newblock
\urldef\tempurl%
\url{https://doi.org/10.1109/FTFC.2013.6577781}
\showDOI{\tempurl}


\bibitem[Nag et~al\mbox{.}(2018)]%
        {Newton}
\bibfield{author}{\bibinfo{person}{Anirban Nag}, \bibinfo{person}{Rajeev
  Balasubramonian}, \bibinfo{person}{Vivek Srikumar}, \bibinfo{person}{Ross
  Walker}, \bibinfo{person}{Ali Shafiee}, \bibinfo{person}{John~Paul Strachan},
  {and} \bibinfo{person}{Naveen Muralimanohar}.}
  \bibinfo{year}{2018}\natexlab{}.
\newblock \showarticletitle{Newton: Gravitating Towards the Physical Limits of
  Crossbar Acceleration}.
\newblock \bibinfo{journal}{\emph{IEEE Micro}} \bibinfo{volume}{38},
  \bibinfo{number}{5} (\bibinfo{year}{2018}), \bibinfo{pages}{41--49}.
\newblock
\urldef\tempurl%
\url{https://doi.org/10.1109/MM.2018.053631140}
\showDOI{\tempurl}


\bibitem[Nagel et~al\mbox{.}(2019)]%
        {data_free_quant}
\bibfield{author}{\bibinfo{person}{Markus Nagel}, \bibinfo{person}{Mart van
  Baalen}, \bibinfo{person}{Tijmen Blankevoort}, {and} \bibinfo{person}{Max
  Welling}.} \bibinfo{year}{2019}\natexlab{}.
\newblock \showarticletitle{Data-Free Quantization Through Weight Equalization
  and Bias Correction}.
\newblock  (\bibinfo{year}{2019}).
\newblock
\urldef\tempurl%
\url{https://doi.org/10.48550/ARXIV.1906.04721}
\showDOI{\tempurl}


\bibitem[O'Halloran and Sarpeshkar(2004)]%
        {sample_and_hold}
\bibfield{author}{\bibinfo{person}{M. O'Halloran} {and} \bibinfo{person}{R.
  Sarpeshkar}.} \bibinfo{year}{2004}\natexlab{}.
\newblock \showarticletitle{A 10-nW 12-bit accurate analog storage cell with
  10-aA leakage}.
\newblock \bibinfo{journal}{\emph{IEEE Journal of Solid-State Circuits}}
  \bibinfo{volume}{39}, \bibinfo{number}{11} (\bibinfo{year}{2004}),
  \bibinfo{pages}{1985--1996}.
\newblock
\urldef\tempurl%
\url{https://doi.org/10.1109/JSSC.2004.835817}
\showDOI{\tempurl}


\bibitem[Okazaki et~al\mbox{.}(2022)]%
        {analog_transformer_1}
\bibfield{author}{\bibinfo{person}{Atsuya Okazaki}, \bibinfo{person}{Pritish
  Narayanan}, \bibinfo{person}{Stefano Ambrogio}, \bibinfo{person}{Kohji
  Hosokawa}, \bibinfo{person}{Hsinyu Tsai}, \bibinfo{person}{Akiyo Nomura},
  \bibinfo{person}{Takeo Yasuda}, \bibinfo{person}{Charles Mackin},
  \bibinfo{person}{Alexander Friz}, \bibinfo{person}{Masatoshi Ishii},
  \bibinfo{person}{Yasuteru Kohda}, \bibinfo{person}{Katie Spoon},
  \bibinfo{person}{An Chen}, \bibinfo{person}{Andrea Fasoli},
  \bibinfo{person}{Malte~J. Rasch}, {and} \bibinfo{person}{Geoffrey~W. Burr}.}
  \bibinfo{year}{2022}\natexlab{}.
\newblock \showarticletitle{Analog-memory-based 14nm Hardware Accelerator for
  Dense Deep Neural Networks including Transformers}. In
  \bibinfo{booktitle}{\emph{2022 IEEE International Symposium on Circuits and
  Systems (ISCAS)}}. \bibinfo{pages}{3319--3323}.
\newblock
\urldef\tempurl%
\url{https://doi.org/10.1109/ISCAS48785.2022.9937292}
\showDOI{\tempurl}


\bibitem[Okumura et~al\mbox{.}(2019)]%
        {ternary_sram_2}
\bibfield{author}{\bibinfo{person}{Shunsuke Okumura}, \bibinfo{person}{Makoto
  Yabuuchi}, \bibinfo{person}{Kenichiro Hijioka}, {and} \bibinfo{person}{Koichi
  Nose}.} \bibinfo{year}{2019}\natexlab{}.
\newblock \showarticletitle{A Ternary Based Bit Scalable, 8.80 TOPS/W CNN
  accelerator with Many-core Processing-in-memory Architecture with 896K
  synapses/mm2}. In \bibinfo{booktitle}{\emph{2019 Symposium on VLSI
  Technology}}. \bibinfo{pages}{C248--C249}.
\newblock
\urldef\tempurl%
\url{https://doi.org/10.23919/VLSIT.2019.8776544}
\showDOI{\tempurl}


\bibitem[Parashar et~al\mbox{.}(2019)]%
        {Timeloop}
\bibfield{author}{\bibinfo{person}{Angshuman Parashar},
  \bibinfo{person}{Priyanka Raina}, \bibinfo{person}{Yakun~Sophia Shao},
  \bibinfo{person}{Yu-Hsin Chen}, \bibinfo{person}{Victor~A. Ying},
  \bibinfo{person}{Anurag Mukkara}, \bibinfo{person}{Rangharajan Venkatesan},
  \bibinfo{person}{Brucek Khailany}, \bibinfo{person}{Stephen~W. Keckler},
  {and} \bibinfo{person}{Joel Emer}.} \bibinfo{year}{2019}\natexlab{}.
\newblock \showarticletitle{Timeloop: A Systematic Approach to DNN Accelerator
  Evaluation}. In \bibinfo{booktitle}{\emph{2019 IEEE International Symposium
  on Performance Analysis of Systems and Software (ISPASS)}}.
  \bibinfo{pages}{304--315}.
\newblock
\urldef\tempurl%
\url{https://doi.org/10.1109/ISPASS.2019.00042}
\showDOI{\tempurl}


\bibitem[Paszke et~al\mbox{.}(2019)]%
        {pytorch}
\bibfield{author}{\bibinfo{person}{Adam Paszke}, \bibinfo{person}{Sam Gross},
  \bibinfo{person}{Francisco Massa}, \bibinfo{person}{Adam Lerer},
  \bibinfo{person}{James Bradbury}, \bibinfo{person}{Gregory Chanan},
  \bibinfo{person}{Trevor Killeen}, \bibinfo{person}{Zeming Lin},
  \bibinfo{person}{Natalia Gimelshein}, \bibinfo{person}{Luca Antiga},
  \bibinfo{person}{Alban Desmaison}, \bibinfo{person}{Andreas Kopf},
  \bibinfo{person}{Edward Yang}, \bibinfo{person}{Zachary DeVito},
  \bibinfo{person}{Martin Raison}, \bibinfo{person}{Alykhan Tejani},
  \bibinfo{person}{Sasank Chilamkurthy}, \bibinfo{person}{Benoit Steiner},
  \bibinfo{person}{Lu Fang}, \bibinfo{person}{Junjie Bai}, {and}
  \bibinfo{person}{Soumith Chintala}.} \bibinfo{year}{2019}\natexlab{}.
\newblock \showarticletitle{PyTorch: An Imperative Style, High-Performance Deep
  Learning Library}.
\newblock In \bibinfo{booktitle}{\emph{Advances in Neural Information
  Processing Systems 32}}. \bibinfo{publisher}{Curran Associates, Inc.},
  \bibinfo{pages}{8024--8035}.
\newblock
\urldef\tempurl%
\url{http://papers.neurips.cc/paper/9015-pytorch-an-imperative-style-high-performance-deep-learning-library.pdf}
\showURL{%
\tempurl}


\bibitem[Patterson et~al\mbox{.}(2021)]%
        {DNN_energy}
\bibfield{author}{\bibinfo{person}{David Patterson}, \bibinfo{person}{Joseph
  Gonzalez}, \bibinfo{person}{Quoc Le}, \bibinfo{person}{Chen Liang},
  \bibinfo{person}{Lluis-Miquel Munguia}, \bibinfo{person}{Daniel Rothchild},
  \bibinfo{person}{David So}, \bibinfo{person}{Maud Texier}, {and}
  \bibinfo{person}{Jeff Dean}.} \bibinfo{year}{2021}\natexlab{}.
\newblock \bibinfo{title}{Carbon Emissions and Large Neural Network Training}.
\newblock
\newblock
\urldef\tempurl%
\url{https://doi.org/10.48550/ARXIV.2104.10350}
\showDOI{\tempurl}


\bibitem[Peng et~al\mbox{.}(2021)]%
        {DNN+NeuroSim}
\bibfield{author}{\bibinfo{person}{Xiaochen Peng}, \bibinfo{person}{Shanshi
  Huang}, \bibinfo{person}{Hongwu Jiang}, \bibinfo{person}{Anni Lu}, {and}
  \bibinfo{person}{Shimeng Yu}.} \bibinfo{year}{2021}\natexlab{}.
\newblock \showarticletitle{DNN+NeuroSim V2.0: An End-to-End Benchmarking
  Framework for Compute-in-Memory Accelerators for On-Chip Training}.
\newblock \bibinfo{journal}{\emph{IEEE Transactions on Computer-Aided Design of
  Integrated Circuits and Systems}} \bibinfo{volume}{40}, \bibinfo{number}{11}
  (\bibinfo{year}{2021}), \bibinfo{pages}{2306--2319}.
\newblock
\urldef\tempurl%
\url{https://doi.org/10.1109/TCAD.2020.3043731}
\showDOI{\tempurl}


\bibitem[Pentecost et~al\mbox{.}(2022)]%
        {NVMExplorer}
\bibfield{author}{\bibinfo{person}{Lillian Pentecost},
  \bibinfo{person}{Alexander Hankin}, \bibinfo{person}{Marco Donato},
  \bibinfo{person}{Mark Hempstead}, \bibinfo{person}{Gu-Yeon Wei}, {and}
  \bibinfo{person}{David Brooks}.} \bibinfo{year}{2022}\natexlab{}.
\newblock \showarticletitle{NVMExplorer: A Framework for Cross-Stack
  Comparisons of Embedded Non-Volatile Memories}. In
  \bibinfo{booktitle}{\emph{2022 IEEE International Symposium on
  High-Performance Computer Architecture (HPCA)}}. \bibinfo{pages}{938--956}.
\newblock
\urldef\tempurl%
\url{https://doi.org/10.1109/HPCA53966.2022.00073}
\showDOI{\tempurl}


\bibitem[Polino et~al\mbox{.}(2018)]%
        {quantization_error_is_gaussian_noise}
\bibfield{author}{\bibinfo{person}{Antonio Polino}, \bibinfo{person}{Razvan
  Pascanu}, {and} \bibinfo{person}{Dan Alistarh}.}
  \bibinfo{year}{2018}\natexlab{}.
\newblock \bibinfo{title}{Model compression via distillation and quantization}.
\newblock
\newblock
\urldef\tempurl%
\url{https://doi.org/10.48550/ARXIV.1802.05668}
\showDOI{\tempurl}


\bibitem[Qiao et~al\mbox{.}(2018)]%
        {AtomLayer}
\bibfield{author}{\bibinfo{person}{Ximing Qiao}, \bibinfo{person}{Xiong Cao},
  \bibinfo{person}{Huanrui Yang}, \bibinfo{person}{Linghao Song}, {and}
  \bibinfo{person}{Hai Li}.} \bibinfo{year}{2018}\natexlab{}.
\newblock \showarticletitle{AtomLayer: A Universal ReRAM-Based CNN Accelerator
  with Atomic Layer Computation}. In \bibinfo{booktitle}{\emph{2018 55th
  ACM/ESDA/IEEE Design Automation Conference (DAC)}}. \bibinfo{pages}{1--6}.
\newblock
\urldef\tempurl%
\url{https://doi.org/10.1109/DAC.2018.8465832}
\showDOI{\tempurl}


\bibitem[Qu et~al\mbox{.}(2021)]%
        {ASBP}
\bibfield{author}{\bibinfo{person}{Songyun Qu}, \bibinfo{person}{Bing Li},
  \bibinfo{person}{Ying Wang}, {and} \bibinfo{person}{Lei Zhang}.}
  \bibinfo{year}{2021}\natexlab{}.
\newblock \showarticletitle{ASBP: Automatic Structured Bit-Pruning for
  RRAM-based NN Accelerator}. In \bibinfo{booktitle}{\emph{2021 58th ACM/IEEE
  Design Automation Conference (DAC)}}. \bibinfo{pages}{745--750}.
\newblock
\urldef\tempurl%
\url{https://doi.org/10.1109/DAC18074.2021.9586105}
\showDOI{\tempurl}


\bibitem[Rajpurkar et~al\mbox{.}(2016)]%
        {SQuAD}
\bibfield{author}{\bibinfo{person}{Pranav Rajpurkar}, \bibinfo{person}{Jian
  Zhang}, \bibinfo{person}{Konstantin Lopyrev}, {and} \bibinfo{person}{Percy
  Liang}.} \bibinfo{year}{2016}\natexlab{}.
\newblock \showarticletitle{{SQ}u{AD}: 100,000+ Questions for Machine
  Comprehension of Text}. In \bibinfo{booktitle}{\emph{Proceedings of the 2016
  Conference on Empirical Methods in Natural Language Processing}}.
  \bibinfo{publisher}{Association for Computational Linguistics},
  \bibinfo{address}{Austin, Texas}, \bibinfo{pages}{2383--2392}.
\newblock
\urldef\tempurl%
\url{https://doi.org/10.18653/v1/D16-1264}
\showDOI{\tempurl}


\bibitem[Ramesh et~al\mbox{.}(2022)]%
        {dalle_proprietary}
\bibfield{author}{\bibinfo{person}{Aditya Ramesh}, \bibinfo{person}{Prafulla
  Dhariwal}, \bibinfo{person}{Alex Nichol}, \bibinfo{person}{Casey Chu}, {and}
  \bibinfo{person}{Mark Chen}.} \bibinfo{year}{2022}\natexlab{}.
\newblock \bibinfo{title}{Hierarchical Text-Conditional Image Generation with
  CLIP Latents}.
\newblock
\newblock
\urldef\tempurl%
\url{https://doi.org/10.48550/ARXIV.2204.06125}
\showDOI{\tempurl}


\bibitem[Rokh et~al\mbox{.}(2022)]%
        {Rokh2022ACS}
\bibfield{author}{\bibinfo{person}{Babak Rokh}, \bibinfo{person}{Ali
  Azarpeyvand}, {and} \bibinfo{person}{Ali~Reza Khanteymoori}.}
  \bibinfo{year}{2022}\natexlab{}.
\newblock \showarticletitle{A Comprehensive Survey on Model Quantization for
  Deep Neural Networks}.
\newblock \bibinfo{journal}{\emph{ArXiv}}  \bibinfo{volume}{abs/2205.07877}
  (\bibinfo{year}{2022}).
\newblock


\bibitem[Saberi et~al\mbox{.}(2011)]%
        {ADC_scaling}
\bibfield{author}{\bibinfo{person}{Mehdi Saberi}, \bibinfo{person}{Reza Lotfi},
  \bibinfo{person}{Khalil Mafinezhad}, {and} \bibinfo{person}{Wouter~A.
  Serdijn}.} \bibinfo{year}{2011}\natexlab{}.
\newblock \showarticletitle{Analysis of Power Consumption and Linearity in
  Capacitive Digital-to-Analog Converters Used in Successive Approximation
  ADCs}.
\newblock \bibinfo{journal}{\emph{IEEE Transactions on Circuits and Systems I:
  Regular Papers}} \bibinfo{volume}{58}, \bibinfo{number}{8}
  (\bibinfo{year}{2011}), \bibinfo{pages}{1736--1748}.
\newblock
\urldef\tempurl%
\url{https://doi.org/10.1109/TCSI.2011.2107214}
\showDOI{\tempurl}


\bibitem[Sandler et~al\mbox{.}(2018)]%
        {MobileNetV2}
\bibfield{author}{\bibinfo{person}{Mark Sandler}, \bibinfo{person}{Andrew
  Howard}, \bibinfo{person}{Menglong Zhu}, \bibinfo{person}{Andrey Zhmoginov},
  {and} \bibinfo{person}{Liang-Chieh Chen}.} \bibinfo{year}{2018}\natexlab{}.
\newblock \showarticletitle{MobileNetV2: Inverted Residuals and Linear
  Bottlenecks}. \bibinfo{pages}{4510--4520}.
\newblock
\urldef\tempurl%
\url{https://doi.org/10.1109/CVPR.2018.00474}
\showDOI{\tempurl}


\bibitem[Shafiee et~al\mbox{.}(2016)]%
        {ISAAC}
\bibfield{author}{\bibinfo{person}{Ali Shafiee}, \bibinfo{person}{Anirban Nag},
  \bibinfo{person}{Naveen Muralimanohar}, \bibinfo{person}{Rajeev
  Balasubramonian}, \bibinfo{person}{John~Paul Strachan}, \bibinfo{person}{Miao
  Hu}, \bibinfo{person}{R.~Stanley Williams}, {and} \bibinfo{person}{Vivek
  Srikumar}.} \bibinfo{year}{2016}\natexlab{}.
\newblock \showarticletitle{ISAAC: A Convolutional Neural Network Accelerator
  with In-Situ Analog Arithmetic in Crossbars}. In
  \bibinfo{booktitle}{\emph{2016 ACM/IEEE 43rd Annual International Symposium
  on Computer Architecture (ISCA)}}. \bibinfo{pages}{14--26}.
\newblock
\urldef\tempurl%
\url{https://doi.org/10.1109/ISCA.2016.12}
\showDOI{\tempurl}


\bibitem[Sinangil et~al\mbox{.}(2021)]%
        {Sinangil}
\bibfield{author}{\bibinfo{person}{Mahmut~E. Sinangil}, \bibinfo{person}{Burak
  Erbagci}, \bibinfo{person}{Rawan Naous}, \bibinfo{person}{Kerem Akarvardar},
  \bibinfo{person}{Dar Sun}, \bibinfo{person}{Win-San Khwa},
  \bibinfo{person}{Hung-Jen Liao}, \bibinfo{person}{Yih Wang}, {and}
  \bibinfo{person}{Jonathan Chang}.} \bibinfo{year}{2021}\natexlab{}.
\newblock \showarticletitle{A 7-nm Compute-in-Memory SRAM Macro Supporting
  Multi-Bit Input, Weight and Output and Achieving 351 TOPS/W and 372.4 GOPS}.
\newblock \bibinfo{journal}{\emph{IEEE Journal of Solid-State Circuits}}
  \bibinfo{volume}{56}, \bibinfo{number}{1} (\bibinfo{year}{2021}),
  \bibinfo{pages}{188--198}.
\newblock
\urldef\tempurl%
\url{https://doi.org/10.1109/JSSC.2020.3031290}
\showDOI{\tempurl}


\bibitem[Song et~al\mbox{.}(2017)]%
        {PipeLayer}
\bibfield{author}{\bibinfo{person}{Linghao Song}, \bibinfo{person}{Xuehai
  Qian}, \bibinfo{person}{Hai Li}, {and} \bibinfo{person}{Yiran Chen}.}
  \bibinfo{year}{2017}\natexlab{}.
\newblock \showarticletitle{PipeLayer: A Pipelined ReRAM-Based Accelerator for
  Deep Learning}. In \bibinfo{booktitle}{\emph{2017 IEEE International
  Symposium on High Performance Computer Architecture (HPCA)}}.
  \bibinfo{pages}{541--552}.
\newblock
\urldef\tempurl%
\url{https://doi.org/10.1109/HPCA.2017.55}
\showDOI{\tempurl}


\bibitem[Song et~al\mbox{.}(2021)]%
        {BRAHMS}
\bibfield{author}{\bibinfo{person}{Tao Song}, \bibinfo{person}{Xiaoming Chen},
  \bibinfo{person}{Xiaoyu Zhang}, {and} \bibinfo{person}{Yinhe Han}.}
  \bibinfo{year}{2021}\natexlab{}.
\newblock \showarticletitle{BRAHMS: Beyond Conventional RRAM-based Neural
  Network Accelerators Using Hybrid Analog Memory System}. In
  \bibinfo{booktitle}{\emph{2021 58th ACM/IEEE Design Automation Conference
  (DAC)}}. \bibinfo{pages}{1033--1038}.
\newblock
\urldef\tempurl%
\url{https://doi.org/10.1109/DAC18074.2021.9586247}
\showDOI{\tempurl}


\bibitem[Spoon et~al\mbox{.}(2021)]%
        {analog_transformer_2}
\bibfield{author}{\bibinfo{person}{Katie Spoon}, \bibinfo{person}{Hsinyu Tsai},
  \bibinfo{person}{An Chen}, \bibinfo{person}{Malte~J. Rasch},
  \bibinfo{person}{Stefano Ambrogio}, \bibinfo{person}{Charles Mackin},
  \bibinfo{person}{Andrea Fasoli}, \bibinfo{person}{Alexander~M. Friz},
  \bibinfo{person}{Pritish Narayanan}, \bibinfo{person}{Milos Stanisavljevic},
  {and} \bibinfo{person}{Geoffrey~W. Burr}.} \bibinfo{year}{2021}\natexlab{}.
\newblock \showarticletitle{Toward Software-Equivalent Accuracy on
  Transformer-Based Deep Neural Networks With Analog Memory Devices}.
\newblock \bibinfo{journal}{\emph{Frontiers in Computational Neuroscience}}
  \bibinfo{volume}{15} (\bibinfo{year}{2021}).
\newblock
\showISSN{1662-5188}
\urldef\tempurl%
\url{https://doi.org/10.3389/fncom.2021.675741}
\showDOI{\tempurl}


\bibitem[Sze et~al\mbox{.}(2020)]%
        {efficient_processing_of_dnns}
\bibfield{author}{\bibinfo{person}{Vivienne Sze}, \bibinfo{person}{Yu-Hsin
  Chen}, \bibinfo{person}{Tien-Ju Yang}, {and} \bibinfo{person}{Joel~S. Emer}.}
  \bibinfo{year}{2020}\natexlab{}.
\newblock \bibinfo{booktitle}{\emph{Efficient Processing of Deep Neural
  Networks}}.
\newblock \bibinfo{publisher}{Springer International Publishing}.
\newblock
\urldef\tempurl%
\url{https://doi.org/10.1007/978-3-031-01766-7}
\showDOI{\tempurl}


\bibitem[Szegedy et~al\mbox{.}(2015)]%
        {GoogLeNet}
\bibfield{author}{\bibinfo{person}{Christian Szegedy}, \bibinfo{person}{Wei
  Liu}, \bibinfo{person}{Yangqing Jia}, \bibinfo{person}{Pierre Sermanet},
  \bibinfo{person}{Scott Reed}, \bibinfo{person}{Dragomir Anguelov},
  \bibinfo{person}{Dumitru Erhan}, \bibinfo{person}{Vincent Vanhoucke}, {and}
  \bibinfo{person}{Andrew Rabinovich}.} \bibinfo{year}{2015}\natexlab{}.
\newblock \showarticletitle{Going deeper with convolutions}. In
  \bibinfo{booktitle}{\emph{2015 IEEE Conference on Computer Vision and Pattern
  Recognition (CVPR)}}. \bibinfo{pages}{1--9}.
\newblock
\urldef\tempurl%
\url{https://doi.org/10.1109/CVPR.2015.7298594}
\showDOI{\tempurl}


\bibitem[Szegedy et~al\mbox{.}(2016)]%
        {InceptionV3}
\bibfield{author}{\bibinfo{person}{Christian Szegedy}, \bibinfo{person}{Vincent
  Vanhoucke}, \bibinfo{person}{Sergey Ioffe}, \bibinfo{person}{Jon Shlens},
  {and} \bibinfo{person}{Zbigniew Wojna}.} \bibinfo{year}{2016}\natexlab{}.
\newblock \showarticletitle{Rethinking the Inception Architecture for Computer
  Vision}. In \bibinfo{booktitle}{\emph{2016 IEEE Conference on Computer Vision
  and Pattern Recognition (CVPR)}}. \bibinfo{pages}{2818--2826}.
\newblock
\urldef\tempurl%
\url{https://doi.org/10.1109/CVPR.2016.308}
\showDOI{\tempurl}


\bibitem[Taigman et~al\mbox{.}(2014)]%
        {deepface_proprietary}
\bibfield{author}{\bibinfo{person}{Yaniv Taigman}, \bibinfo{person}{Ming Yang},
  \bibinfo{person}{Marc'Aurelio Ranzato}, {and} \bibinfo{person}{Lior Wolf}.}
  \bibinfo{year}{2014}\natexlab{}.
\newblock \showarticletitle{DeepFace: Closing the Gap to Human-Level
  Performance in Face Verification}. In \bibinfo{booktitle}{\emph{2014 IEEE
  Conference on Computer Vision and Pattern Recognition}}.
  \bibinfo{pages}{1701--1708}.
\newblock
\urldef\tempurl%
\url{https://doi.org/10.1109/CVPR.2014.220}
\showDOI{\tempurl}


\bibitem[Tu et~al\mbox{.}(2018)]%
        {eDRAM_refresh}
\bibfield{author}{\bibinfo{person}{Fengbin Tu}, \bibinfo{person}{Weiwei Wu},
  \bibinfo{person}{Shouyi Yin}, \bibinfo{person}{Leibo Liu}, {and}
  \bibinfo{person}{Shaojun Wei}.} \bibinfo{year}{2018}\natexlab{}.
\newblock \showarticletitle{RANA: Towards Efficient Neural Acceleration with
  Refresh-Optimized Embedded DRAM}.
\newblock \bibinfo{journal}{\emph{2018 ACM/IEEE 45th Annual International
  Symposium on Computer Architecture (ISCA)}} (\bibinfo{year}{2018}),
  \bibinfo{pages}{340--352}.
\newblock


\bibitem[Vaswani et~al\mbox{.}(2017)]%
        {transformer}
\bibfield{author}{\bibinfo{person}{Ashish Vaswani}, \bibinfo{person}{Noam
  Shazeer}, \bibinfo{person}{Niki Parmar}, \bibinfo{person}{Jakob Uszkoreit},
  \bibinfo{person}{Llion Jones}, \bibinfo{person}{Aidan~N Gomez},
  \bibinfo{person}{\L~ukasz Kaiser}, {and} \bibinfo{person}{Illia Polosukhin}.}
  \bibinfo{year}{2017}\natexlab{}.
\newblock \showarticletitle{Attention is All you Need}. In
  \bibinfo{booktitle}{\emph{Advances in Neural Information Processing
  Systems}}, \bibfield{editor}{\bibinfo{person}{I.~Guyon},
  \bibinfo{person}{U.~Von Luxburg}, \bibinfo{person}{S.~Bengio},
  \bibinfo{person}{H.~Wallach}, \bibinfo{person}{R.~Fergus},
  \bibinfo{person}{S.~Vishwanathan}, {and} \bibinfo{person}{R.~Garnett}}
  (Eds.), Vol.~\bibinfo{volume}{30}. \bibinfo{publisher}{Curran Associates,
  Inc.}
\newblock
\urldef\tempurl%
\url{https://proceedings.neurips.cc/paper/2017/file/3f5ee243547dee91fbd053c1c4a845aa-Paper.pdf}
\showURL{%
\tempurl}


\bibitem[Verhelst and Murmann(2012)]%
        {ADC_Scaling_Murmann}
\bibfield{author}{\bibinfo{person}{Marian Verhelst} {and}
  \bibinfo{person}{Boris Murmann}.} \bibinfo{year}{2012}\natexlab{}.
\newblock \showarticletitle{Area scaling analysis of CMOS ADCs}.
\newblock \bibinfo{journal}{\emph{Electronics Letters}}  \bibinfo{volume}{48}
  (\bibinfo{year}{2012}), \bibinfo{pages}{314--315}.
\newblock


\bibitem[Wang et~al\mbox{.}(2010)]%
        {4F_1T1R}
\bibfield{author}{\bibinfo{person}{Ching-Hua Wang}, \bibinfo{person}{Yi-Hung
  Tsai}, \bibinfo{person}{Kai-Chun Lin}, \bibinfo{person}{Meng-Fan Chang},
  \bibinfo{person}{Ya-Chin King}, \bibinfo{person}{Chrong-Jung Lin},
  \bibinfo{person}{Shyh-Shyuan Sheu}, \bibinfo{person}{Yu-Sheng Chen},
  \bibinfo{person}{Heng-Yuan Lee}, \bibinfo{person}{Frederick~T. Chen}, {and}
  \bibinfo{person}{Ming-Jinn Tsai}.} \bibinfo{year}{2010}\natexlab{}.
\newblock \showarticletitle{Three-dimensional 4F2 ReRAM cell with CMOS logic
  compatible process}. In \bibinfo{booktitle}{\emph{2010 International Electron
  Devices Meeting}}. \bibinfo{pages}{29.6.1--29.6.4}.
\newblock
\urldef\tempurl%
\url{https://doi.org/10.1109/IEDM.2010.5703446}
\showDOI{\tempurl}


\bibitem[Wang et~al\mbox{.}(2021)]%
        {2T2R}
\bibfield{author}{\bibinfo{person}{Linfang Wang}, \bibinfo{person}{Wang Ye},
  \bibinfo{person}{Chunmeng Dou}, \bibinfo{person}{Xin Si},
  \bibinfo{person}{Xiaoxin Xu}, \bibinfo{person}{Jing Liu},
  \bibinfo{person}{Dashan Shang}, \bibinfo{person}{Jianfeng Gao},
  \bibinfo{person}{Feng Zhang}, \bibinfo{person}{Yongpan Liu},
  \bibinfo{person}{Meng-Fan Chang}, {and} \bibinfo{person}{Qi Liu}.}
  \bibinfo{year}{2021}\natexlab{}.
\newblock \showarticletitle{Efficient and Robust Nonvolatile
  Computing-In-Memory Based on Voltage Division in 2T2R RRAM With
  Input-Dependent Sensing Control}.
\newblock \bibinfo{journal}{\emph{IEEE Transactions on Circuits and Systems II:
  Express Briefs}} \bibinfo{volume}{68}, \bibinfo{number}{5}
  (\bibinfo{year}{2021}), \bibinfo{pages}{1640--1644}.
\newblock
\urldef\tempurl%
\url{https://doi.org/10.1109/TCSII.2021.3067385}
\showDOI{\tempurl}


\bibitem[Wikipedia(2022)]%
        {wiki:Iron_law_of_processor_performance}
\bibfield{author}{\bibinfo{person}{Wikipedia}.}
  \bibinfo{year}{2022}\natexlab{}.
\newblock \bibinfo{title}{{Iron law of processor performance} ---
  {W}ikipedia{,} The Free Encyclopedia}.
\newblock
  \bibinfo{howpublished}{\url{http://en.wikipedia.org/w/index.php?title=Iron\%20law\%20of\%20processor\%20performance&oldid=1112639388}}.
\newblock
\newblock
\shownote{[Online; accessed 22-November-2022]}.


\bibitem[Wu et~al\mbox{.}(2020a)]%
        {Wu2020IntegerQF}
\bibfield{author}{\bibinfo{person}{Hao Wu}, \bibinfo{person}{Patrick Judd},
  \bibinfo{person}{Xiaojie Zhang}, \bibinfo{person}{Mikhail Isaev}, {and}
  \bibinfo{person}{Paulius Micikevicius}.} \bibinfo{year}{2020}\natexlab{a}.
\newblock \showarticletitle{Integer Quantization for Deep Learning Inference:
  Principles and Empirical Evaluation}.
\newblock \bibinfo{journal}{\emph{ArXiv}}  \bibinfo{volume}{abs/2004.09602}
  (\bibinfo{year}{2020}).
\newblock


\bibitem[Wu et~al\mbox{.}(2020b)]%
        {qdqbert}
\bibfield{author}{\bibinfo{person}{Hao Wu}, \bibinfo{person}{Patrick Judd},
  \bibinfo{person}{Xiaojie Zhang}, \bibinfo{person}{Mikhail Isaev}, {and}
  \bibinfo{person}{Paulius Micikevicius}.} \bibinfo{year}{2020}\natexlab{b}.
\newblock \bibinfo{title}{Integer Quantization for Deep Learning Inference:
  Principles and Empirical Evaluation}.
\newblock
\newblock


\bibitem[Wu et~al\mbox{.}(2019)]%
        {accelergy}
\bibfield{author}{\bibinfo{person}{Yannan~Nellie Wu}, \bibinfo{person}{Joel~S.
  Emer}, {and} \bibinfo{person}{Vivienne Sze}.}
  \bibinfo{year}{2019}\natexlab{}.
\newblock \showarticletitle{Accelergy: An Architecture-Level Energy Estimation
  Methodology for Accelerator Designs}. In \bibinfo{booktitle}{\emph{2019
  IEEE/ACM International Conference on Computer-Aided Design (ICCAD)}}.
  \bibinfo{pages}{1--8}.
\newblock
\urldef\tempurl%
\url{https://doi.org/10.1109/ICCAD45719.2019.8942149}
\showDOI{\tempurl}


\bibitem[Wu et~al\mbox{.}(2020c)]%
        {accelergy_pim}
\bibfield{author}{\bibinfo{person}{Yannan~Nellie Wu}, \bibinfo{person}{Vivienne
  Sze}, {and} \bibinfo{person}{Joel~S. Emer}.}
  \bibinfo{year}{2020}\natexlab{c}.
\newblock \showarticletitle{An Architecture-Level Energy and Area Estimator for
  Processing-In-Memory Accelerator Designs}. In \bibinfo{booktitle}{\emph{2020
  IEEE International Symposium on Performance Analysis of Systems and Software
  (ISPASS)}}. \bibinfo{pages}{116--118}.
\newblock
\urldef\tempurl%
\url{https://doi.org/10.1109/ISPASS48437.2020.00024}
\showDOI{\tempurl}


\bibitem[Xiao et~al\mbox{.}(2021)]%
        {fidelity_encoding_exploration}
\bibfield{author}{\bibinfo{person}{T.~Patrick Xiao}, \bibinfo{person}{Ben
  Feinberg}, \bibinfo{person}{Christopher~H. Bennett},
  \bibinfo{person}{Venkatraman Prabhakar}, \bibinfo{person}{Prashant Saxena},
  \bibinfo{person}{Vineet Agrawal}, \bibinfo{person}{Sapan Agarwal}, {and}
  \bibinfo{person}{Matthew~J. Marinella}.} \bibinfo{year}{2021}\natexlab{}.
\newblock \showarticletitle{On the Accuracy of Analog Neural Network Inference
  Accelerators [Feature]}.
\newblock \bibinfo{journal}{\emph{IEEE Circuits and Systems Magazine}}
  \bibinfo{volume}{22} (\bibinfo{year}{2021}), \bibinfo{pages}{26--48}.
\newblock


\bibitem[Yang et~al\mbox{.}(2013)]%
        {stacked_MP_2}
\bibfield{author}{\bibinfo{person}{J.~Joshua Yang}, \bibinfo{person}{Dmitri~B.
  Strukov}, {and} \bibinfo{person}{Duncan~R. Stewart}.}
  \bibinfo{year}{2013}\natexlab{}.
\newblock \showarticletitle{Memristive devices for computing}.
\newblock \bibinfo{journal}{\emph{Nature Nanotechnology}} \bibinfo{volume}{8},
  \bibinfo{number}{1} (\bibinfo{date}{01 Jan} \bibinfo{year}{2013}),
  \bibinfo{pages}{13--24}.
\newblock
\showISSN{1748-3395}
\urldef\tempurl%
\url{https://doi.org/10.1038/nnano.2012.240}
\showDOI{\tempurl}


\bibitem[Yang et~al\mbox{.}(2019)]%
        {SRE}
\bibfield{author}{\bibinfo{person}{Tzu-Hsien Yang}, \bibinfo{person}{Hsiang-Yun
  Cheng}, \bibinfo{person}{Chia-Lin Yang}, \bibinfo{person}{I-Ching Tseng},
  \bibinfo{person}{Han-Wen Hu}, \bibinfo{person}{Hung-Sheng Chang}, {and}
  \bibinfo{person}{Hsiang-Pang Li}.} \bibinfo{year}{2019}\natexlab{}.
\newblock \showarticletitle{Sparse ReRAM Engine: Joint Exploration of
  Activation and Weight Sparsity in Compressed Neural Networks}. In
  \bibinfo{booktitle}{\emph{2019 ACM/IEEE 46th Annual International Symposium
  on Computer Architecture (ISCA)}}. \bibinfo{pages}{236--249}.
\newblock


\bibitem[Yang and Sze(2019)]%
        {noisy_small_dnns}
\bibfield{author}{\bibinfo{person}{Tien-Ju Yang} {and}
  \bibinfo{person}{Vivienne Sze}.} \bibinfo{year}{2019}\natexlab{}.
\newblock \showarticletitle{Design Considerations for Efficient Deep Neural
  Networks on Processing-in-Memory Accelerators}.
  \bibinfo{pages}{22.1.1--22.1.4}.
\newblock
\urldef\tempurl%
\url{https://doi.org/10.1109/IEDM19573.2019.8993662}
\showDOI{\tempurl}


\bibitem[Yazdanbakhsh et~al\mbox{.}(2022)]%
        {analog_transformer_3}
\bibfield{author}{\bibinfo{person}{Amir Yazdanbakhsh}, \bibinfo{person}{Ashkan
  Moradifirouzabadi}, \bibinfo{person}{Zheng Li}, {and} \bibinfo{person}{Mingu
  Kang}.} \bibinfo{year}{2022}\natexlab{}.
\newblock \bibinfo{title}{Sparse Attention Acceleration with Synergistic
  In-Memory Pruning and On-Chip Recomputation}.
\newblock
\newblock
\urldef\tempurl%
\url{https://doi.org/10.48550/ARXIV.2209.00606}
\showDOI{\tempurl}


\bibitem[Yin et~al\mbox{.}(2020)]%
        {ternary_sram}
\bibfield{author}{\bibinfo{person}{Shihui Yin}, \bibinfo{person}{Zhewei Jiang},
  \bibinfo{person}{Jae-Sun Seo}, {and} \bibinfo{person}{Mingoo Seok}.}
  \bibinfo{year}{2020}\natexlab{}.
\newblock \showarticletitle{XNOR-SRAM: In-Memory Computing SRAM Macro for
  Binary/Ternary Deep Neural Networks}.
\newblock \bibinfo{journal}{\emph{IEEE Journal of Solid-State Circuits}}
  \bibinfo{volume}{55}, \bibinfo{number}{6} (\bibinfo{year}{2020}),
  \bibinfo{pages}{1733--1743}.
\newblock
\urldef\tempurl%
\url{https://doi.org/10.1109/JSSC.2019.2963616}
\showDOI{\tempurl}


\bibitem[Yuan et~al\mbox{.}(2021a)]%
        {tinyadc}
\bibfield{author}{\bibinfo{person}{Geng Yuan}, \bibinfo{person}{Payman Behnam},
  \bibinfo{person}{Yuxuan Cai}, \bibinfo{person}{Ali Shafiee},
  \bibinfo{person}{Jingyan Fu}, \bibinfo{person}{Zhiheng Liao},
  \bibinfo{person}{Zhengang Li}, \bibinfo{person}{Xiaolong Ma},
  \bibinfo{person}{Jieren Deng}, \bibinfo{person}{Jinhui Wang},
  \bibinfo{person}{Mahdi Bojnordi}, \bibinfo{person}{Yanzhi Wang}, {and}
  \bibinfo{person}{Caiwen Ding}.} \bibinfo{year}{2021}\natexlab{a}.
\newblock \showarticletitle{TinyADC: Peripheral Circuit-aware Weight Pruning
  Framework for Mixed-signal DNN Accelerators}. In
  \bibinfo{booktitle}{\emph{2021 Design, Automation And Test in Europe
  Conference And Exhibition (DATE)}}. \bibinfo{pages}{926--931}.
\newblock
\urldef\tempurl%
\url{https://doi.org/10.23919/DATE51398.2021.9474235}
\showDOI{\tempurl}


\bibitem[Yuan et~al\mbox{.}(2021b)]%
        {FORMS}
\bibfield{author}{\bibinfo{person}{Geng Yuan}, \bibinfo{person}{Payman Behnam},
  \bibinfo{person}{Zhengang Li}, \bibinfo{person}{Ali Shafiee},
  \bibinfo{person}{Sheng Lin}, \bibinfo{person}{Xiaolong Ma},
  \bibinfo{person}{Hang Liu}, \bibinfo{person}{Xuehai Qian},
  \bibinfo{person}{Mahdi~Nazm Bojnordi}, \bibinfo{person}{Yanzhi Wang}, {and}
  \bibinfo{person}{Caiwen Ding}.} \bibinfo{year}{2021}\natexlab{b}.
\newblock \showarticletitle{FORMS: Fine-grained Polarized ReRAM-based In-situ
  Computation for Mixed-signal DNN Accelerator}. In
  \bibinfo{booktitle}{\emph{2021 ACM/IEEE 48th Annual International Symposium
  on Computer Architecture (ISCA)}}. \bibinfo{pages}{265--278}.
\newblock
\urldef\tempurl%
\url{https://doi.org/10.1109/ISCA52012.2021.00029}
\showDOI{\tempurl}


\bibitem[Yue et~al\mbox{.}(2019)]%
        {1T2R_Aeris}
\bibfield{author}{\bibinfo{person}{Jinshan Yue}, \bibinfo{person}{Yongpan Liu},
  \bibinfo{person}{Fang Su}, \bibinfo{person}{Shuangchen Li},
  \bibinfo{person}{Zhe Yuan}, \bibinfo{person}{Zhibo Wang},
  \bibinfo{person}{Wenyu Sun}, \bibinfo{person}{Xueqing Li}, {and}
  \bibinfo{person}{Huazhong Yang}.} \bibinfo{year}{2019}\natexlab{}.
\newblock \showarticletitle{AERIS: Area/Energy-Efficient 1T2R ReRAM Based
  Processing-in-Memory Neural Network System-on-a-Chip}. In
  \bibinfo{booktitle}{\emph{Proceedings of the 24th Asia and South Pacific
  Design Automation Conference}} (Tokyo, Japan) \emph{(\bibinfo{series}{ASPDAC
  '19})}. \bibinfo{publisher}{Association for Computing Machinery},
  \bibinfo{address}{New York, NY, USA}, \bibinfo{pages}{146–151}.
\newblock
\showISBNx{9781450360074}
\urldef\tempurl%
\url{https://doi.org/10.1145/3287624.3287635}
\showDOI{\tempurl}


\bibitem[Zhao et~al\mbox{.}(2020)]%
        {Zhao2020LinearSQ}
\bibfield{author}{\bibinfo{person}{Xiandong Zhao}, \bibinfo{person}{Ying Wang},
  \bibinfo{person}{Xuyi Cai}, \bibinfo{person}{Chuanming Liu}, {and}
  \bibinfo{person}{Lei Zhang}.} \bibinfo{year}{2020}\natexlab{}.
\newblock \showarticletitle{Linear Symmetric Quantization of Neural Networks
  for Low-precision Integer Hardware}. In \bibinfo{booktitle}{\emph{ICLR}}.
\newblock


\end{thebibliography}
\balance

\end{document}